\documentclass[journal]{IEEEtran}
\usepackage{amsfonts,amsmath,amssymb}
\usepackage{hyperref}
\usepackage{subcaption}
\usepackage{fancyhdr}
\usepackage[export]{adjustbox}

\usepackage{float}
\usepackage[longtable]{multirow}

\usepackage{graphicx}
\usepackage{url}
\usepackage{placeins} % preamble
\usepackage{caption}
\usepackage{cite}

\usepackage{array}

\begin{document}
		\title{Experimental Acquisition and Verification of Spectral Signatures of Dynamic Bifurcations}

		
		\author{Suvradip~Maity, Debajyoti~Guha,
            and~Soumitro~Banerjee% <-this % stops a space
			\thanks{The authors are with the Department of Physical Sciences, Indian Institute of Science Education and Research (IISER) Kolkata, India (e-mails:
sm21ms141@iiserkol.ac.in; dg19rs013@iiserkol.ac.in; soumitro@iiserkol.ac.in). D. Guha is also affiliated with the Department of Physics, School of Basic Sciences, Swami Vivekananda University (Barrackpore), India}
            \thanks{All correspondence to be made with D. Guha. e-mail: drdebajyotiguha@gmail.com}}
	\maketitle
	
    
\begin{abstract}
		Spectral bifurcation diagrams (SBDs) have recently emerged as an efficient tool for identifying dynamical transitions in nonlinear systems through frequency-domain analysis. Previous studies have been limited to numerical investigations, and the experimental realization of SBDs has remained unexplored. In this work, we develop an automated framework using analog electronic circuits and data acquisition (DAQ) systems to obtain SBDs from real-time measurements. The method enables controlled parameter variation and simultaneous acquisition of time-series data for spectral analysis. Using this approach, we experimentally capture characteristic spectral signatures of dynamical bifurcations, such as period-doubling, quasiperiodicity (two- and three-frequency), and torus length-doubling. The experimental results show strong qualitative agreement with the numerical predictions, despite noise and parameter mismatches. This study establishes SBD as an effective tool for the experimental analysis of nonlinear dynamical systems.
\end{abstract}

\begin{IEEEkeywords}
spectral bifurcation diagram, bifurcation analysis, quasiperiodicity, torus-doubling, FFT, DAQ, nonlinear circuits.
\end{IEEEkeywords}
	% \begin{IEEEkeywords}
	% 	FFT, Quasiperiodicity, Torus doubling
	% \end{IEEEkeywords}
	% \fi
	% \IEEEpeerreviewmaketitle

\section{Introduction}

Bifurcation theory plays a pivotal role in understanding the mechanisms underlying transitions between qualitatively distinct behaviors in nonlinear systems. As a system parameter varies, bifurcations can lead to a wide variety of phenomena such as period-doubling, quasiperiodicity, and chaos~\cite{may1987chaos,YO12,chua1993universal1, chaos1993universal2,MLC93,matsumoto1987chaos}. Traditionally, bifurcation analysis relies on tools such as one-parameter and two-parameter bifurcation diagrams~\cite{GU13,Ku08,ricco2016circuit,kawakami1984bifurcation}, Lyapunov exponents~\cite{WO85,YO13,PA93}, and so on. 

Spectral analysis of time-series data can also reveal important aspects of bifurcations, such as abrupt changes in spectral content. It has long been known that Hopf bifurcations in continuous systems and Neimark-Sacker bifurcations in discrete systems give rise to new frequency components. Furthermore, a period-doubling bifurcation gives rise to a new frequency component at half the fundamental frequency existing before the bifurcation~\cite{ST15}.  

Although such phenomena have long been recognized, systematic attempts to study the evolution of spectral components as the system parameters vary are relatively recent. Through the efforts of a few sparse groups around the world, the `spectral bifurcation diagram' has evolved into a computationally efficient tool for bifurcation analysis~\cite{SBD2,SBD3,borkowski2015fft,karimov2021bifurcation,orrell2003visualizing,zandi2022fft}.
Recently, using such spectral bifurcation diagrams, Guha and Banerjee~\cite{GU24} systematically investigated the spectral signatures of various static and dynamic bifurcations and identified distinct spectral patterns for each bifurcation.

In this paper, we report an experimental methodology for directly obtaining spectral bifurcation diagrams from nonlinear circuits and present an experimental verification of spectral behavior previously observed only numerically. The method centers on an automated variation of the control parameter via a software-driven interface while simultaneously storing real-time data for spectral analysis. To assess the robustness of our method, we choose systems that exhibit dynamic bifurcations involving multiple frequencies, namely the secondary Hopf or torus bifurcation, the torus-doubling bifurcation, and the third Hopf (or the 2-Torus-to-3-Torus) bifurcation. 

The paper is organized as follows. In Sec.~\ref{intro}, we briefly reintroduce the procedure to obtain the spectral bifurcation diagram of continuous-time systems numerically as reported in~\cite{GU24}. We then elaborate on the experimental methodology developed to obtain such diagrams. In Sec.~\ref{illus}, we illustrate the methodology considering an example system showing period-doubling bifurcation. We then proceed to apply the method to analyze various dynamic bifurcations in Sec.~\ref{exp_analysis}. We conclude by discussing the results and future applications of our method in Sec.~\ref{conc}.

\section{The spectral bifurcation diagram\label{intro}}

A spectral bifurcation diagram is a panoramic view of how a system's peak frequencies change as a parameter is varied. It captures many dynamic transitions that are absent or ambiguous in a regular bifurcation diagram, as shown by Guha and Banerjee~\cite{GU24}. We shall discuss the methods for numerically and experimentally generating such a diagram for continuous-time systems. One may refer to~\cite{GU24} for a detailed study of spectral signatures in both continuous-time and discrete-time systems.

\subsection{Numerical method}

The first step is to choose a parameter value within the parameter range under consideration, along with a suitable set of initial conditions. For a dynamic system (oscillatory, chaotic, or quasiperiodic), we then solve the given set of differential equations over a sufficiently long time ($t>>100T$, where $T$ is the highest time period of the system). After removing the transients  the last $2^{17}\!-\!2^{24}$ data points are passed through an algorithm incorporating MATLAB's \emph{FFT} function~\cite{MA26a} with a sampling rate at least twice the highest frequency in the system, thus satisfying the Nyquist-Shannon sampling theorem~\cite{NY28,SH48}. We then use MATLAB's \emph{findpeaks} function~\cite{MA26b} to identify the significant frequency components (in Hz) and their respective powers (in dB).

This procedure is repeated over the entire parameter range, in a path-following manner, to avoid attractor switching in the presence of multiple coexisting attractors. Finally, we plot the parameter along the $x$-axis and the frequencies on the $y$-axis, using relative power as a color variable. What we thus obtain is the numerically obtained spectral bifurcation diagram (which we refer to as SBD, hereafter). 

\subsection{Experimental method}

	\begin{figure}[ht]
		\centering		\includegraphics[height=0.2\textwidth,width=0.35\textwidth]{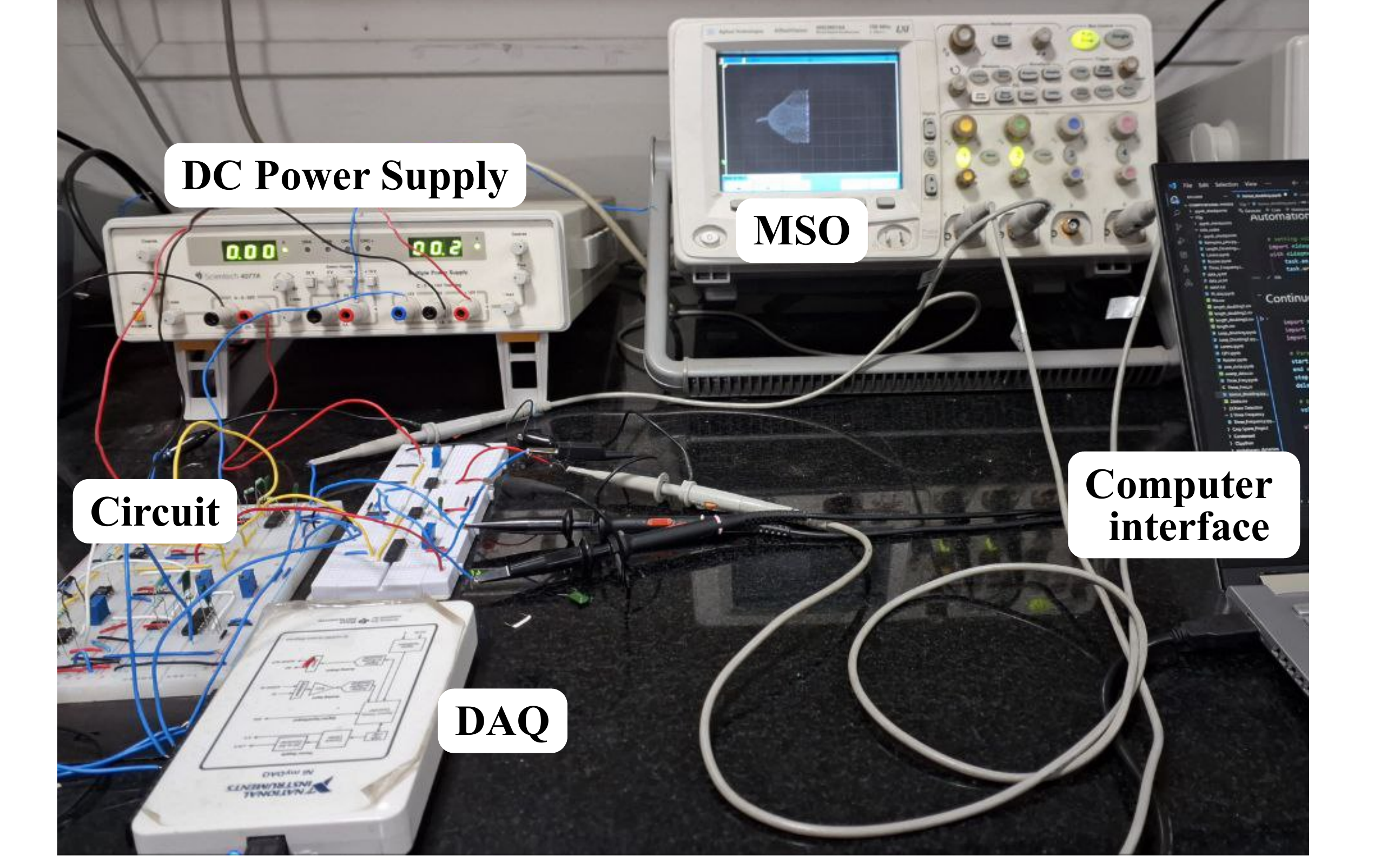}
		\caption{The complete experimental set-up showing the DC power-supply, the breadboard circuitry, the DAQ, and the oscilloscope. The MSO (Mixed Signal Oscilloscope) used is the MSO6014A (Agilent), the DC power supply is the Scientech 4077A, and data acquisition is performed with the NI myDAQ (NI).}
		\label{exp_setup}
	\end{figure}

The experimental set-up is shown in Fig.~\ref{exp_setup}. For our work, we have used nonlinear circuits built from resistors, capacitors, diodes, op-amps, analog multipliers, etc. However, the method is general and can be used on any physical system. The only constraint is that the system's control parameter must be a DC voltage. In our work, we obtain the control parameter using NI myDAQ~\cite{NI26a}. 

As the system runs, the DAQ's input channel automatically collects real-time time-series data for the state variables corresponding to a given parameter value. The sampling rate for this was set to $200 \times 10^3$ samples/s for better representation of the original signal~\cite{NI26b}. Skipping $10^5$ data points for transients, the last $2^{17}$ data points are automatically stored in a CSV file through a Python program. Parameter variation is automated using Python's \emph{nidaqmx} package, and the entire process is repeated for each parameter step in the desired range.

This experimentally obtained data is then run through the same program we used to generate the numerical SBD, which incorporates MATLAB's \emph{FFT} and \emph{findpeaks} functions to filter out the prominent peaks. Finally, plotting the parameter along the $x$-axis, the frequency (in Hz) on the $y$-axis, and the relative power (in dB) as the color variable yields the experimentally obtained spectral bifurcation diagram. 

\subsubsection*{Regular bifurcation diagram}

Due to parameter mismatches, we do not expect the experimental and numerical SBD to be identical. To better compare the dynamic transitions of the experimental systems, we also generate a one-parameter bifurcation diagram experimentally (which we call the `regular' bifurcation diagram to distinguish it from the SBD). 

\begin{figure}[ht]
    \centering
    \includegraphics[width=0.9\linewidth]{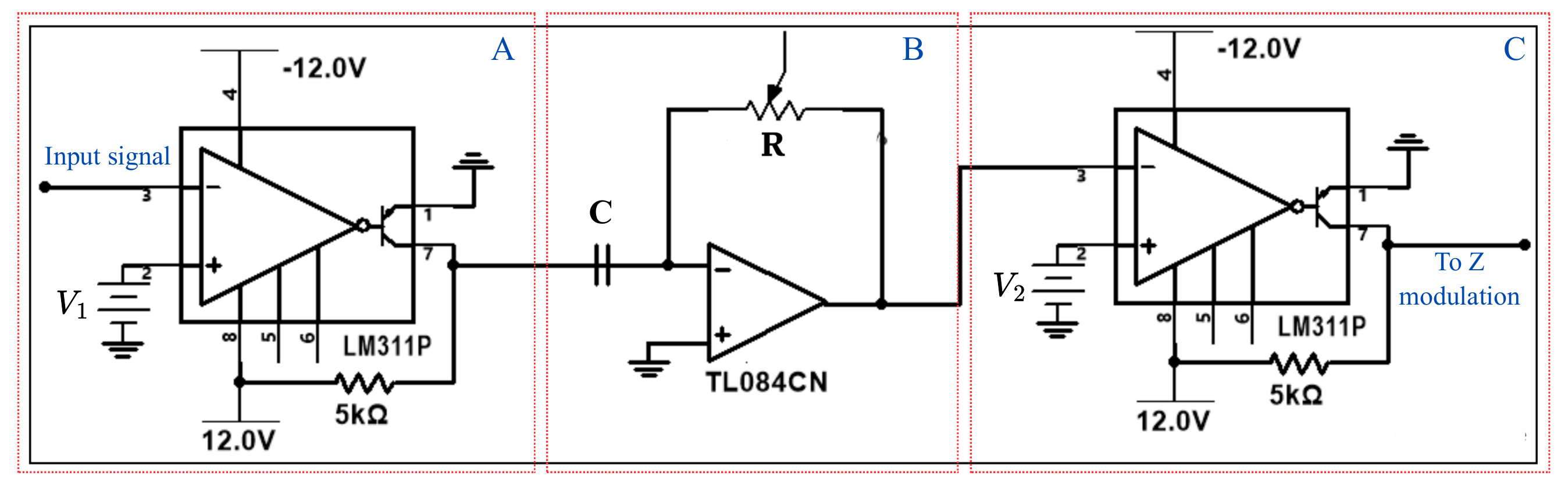}
    \caption{Circuit for obtaining the Poincar\'e section. A and C are comparator blocks, and B is a differentiator block. The reference voltages $\text{V}_1$ and $\text{V}_2$ are suitable voltage values applied to obtain the Poincar\'e section. To enhance differentiation, R and C are chosen such that $RC \ll T'$, where $T'$ is the smallest time period of the input signal. LM311P IC is used as the comparator, and TL084CN IC as the op-amp.}
    \label{Poinc}
\end{figure}

To obtain the regular bifurcation diagram, we construct an additional circuit that is able to obtain a Poincaré cross-section as shown in Fig.~\ref{Poinc}~\cite{SE20}. A state variable, say $x$, is fed into a voltage comparator (LM311P) along with a reference voltage $V_1$ that defines the position of the Poincar\'e plane. This produces a square wave of $\pm 12~\text{V}$ that corresponds to whenever $x$ crosses $V_1$. This is fed into a differentiator circuit, producing an impulse waveform, which is then passed through another comparator to filter out the negative spikes. The final output is fed to the oscilloscope's Z-axis modulation channel. This causes the oscilloscope to display the other state variables only at the instants when $x$ crosses the reference voltage, thus effectively acting as a Poincar\'e cross-section circuit. 

The varying control parameter is read by the X-channel of the oscilloscope, while the sampled state variable is fed to the Y-channel. The experimental bifurcation diagram appears directly on the oscilloscope with the help of the \emph{persistent} functionality of MSO6014A~\cite{MSO}.

\section{Illustrative example} \label{illus}

We now demonstrate the simultaneous acquisition of data for experimental SBD and the generation of a regular bifurcation diagram through automated parameter variation. We begin with a dynamic bifurcation involving only one independent frequency. To demonstrate this, we choose the R\"ossler system~(\ref{rosseqn}), which is well known for showing a period-doubling cascade leading to chaos. In the equations that follow, overdot indicates differentiation with respect to time. 
    \begin{equation}        \label{rosseqn}
        \begin{split}
             \dot{x'} &= - y' - z' \\
             \dot{y'} &= x' + ay' \\
             \dot{z'} &= b + z'(x'-c)
         \end{split}
    \end{equation}

\begin{figure}[t]
    \centering
    \includegraphics[width=0.95\linewidth]{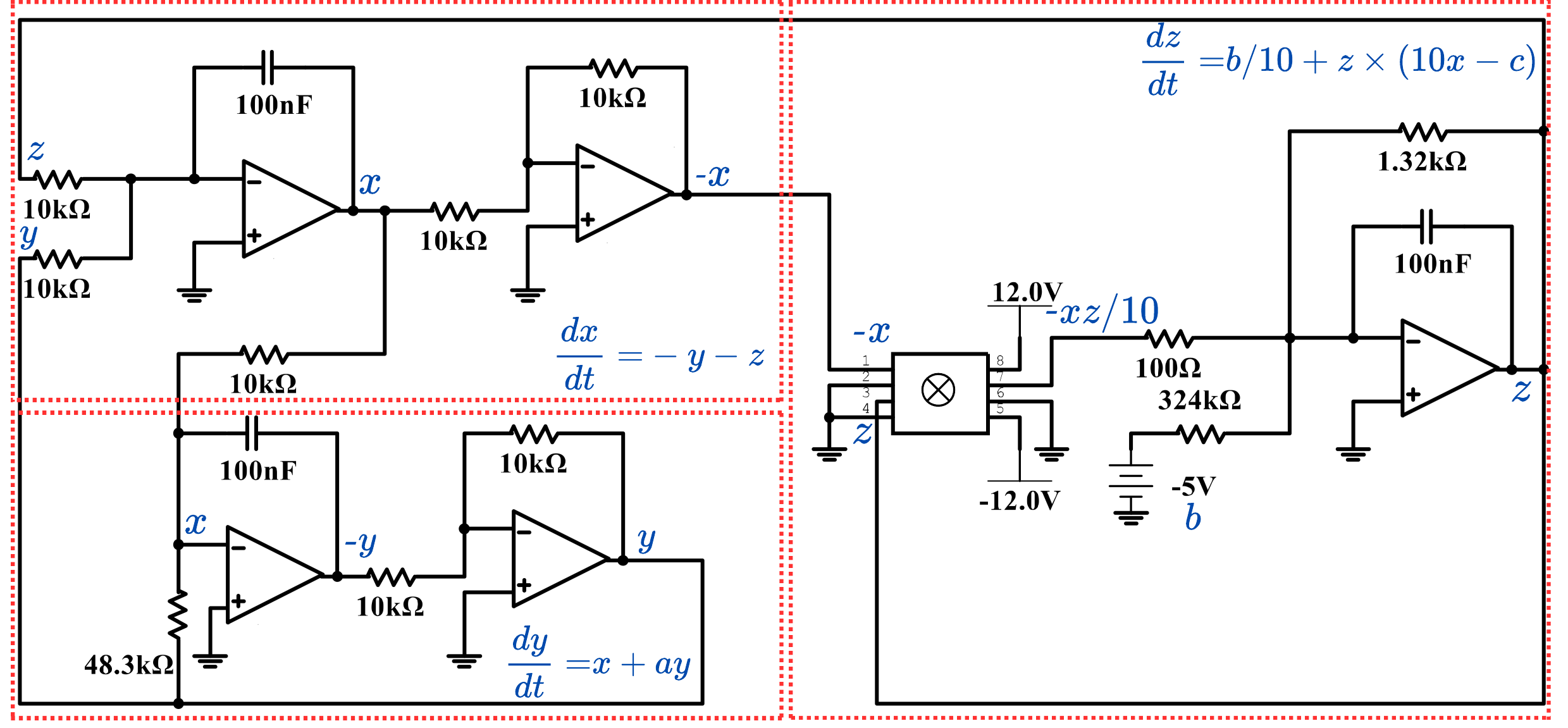}
    \caption{Electronic circuit implementation of the rescaled R\"ossler system given by (\ref{rosseqn2}). The control parameter $b$ is varied using the DAQ output. TL084CN and AD633JN are used as an op-amp and a multiplier (denoted by $\otimes$), respectively.}
    \label{R_circuit}
\end{figure}

The numerical solution shows that the amplitudes of the state variables may exceed $\pm 12$~V, which exceeds the op-amps' saturation voltage. So we scale (\ref{rosseqn}) down by a factor of $10$ (value chosen so that the scaled voltages remain within a detectable regime) and replace the variables (e.g., $x = \tfrac{x'}{10}$) to obtain the rescaled equations (\ref{rosseqn2}). The circuit that simulates (\ref{rosseqn2}) is given in Fig.~\ref{R_circuit}.
	\begin{equation}        \label{rosseqn2}
        \begin{split}
             \dot{x} &= - y - z \\
             \dot{y} &= x + ay \\
             \dot{z} &= \dfrac{b}{10} + z(10x-c)
         \end{split}
    \end{equation}

\subsection{Regular bifurcation diagram}
\begin{figure}[ht]
		\centering
		\begin{subfigure}{\columnwidth}
        \centering
			\includegraphics[height=0.45\textwidth,width=0.7\textwidth]{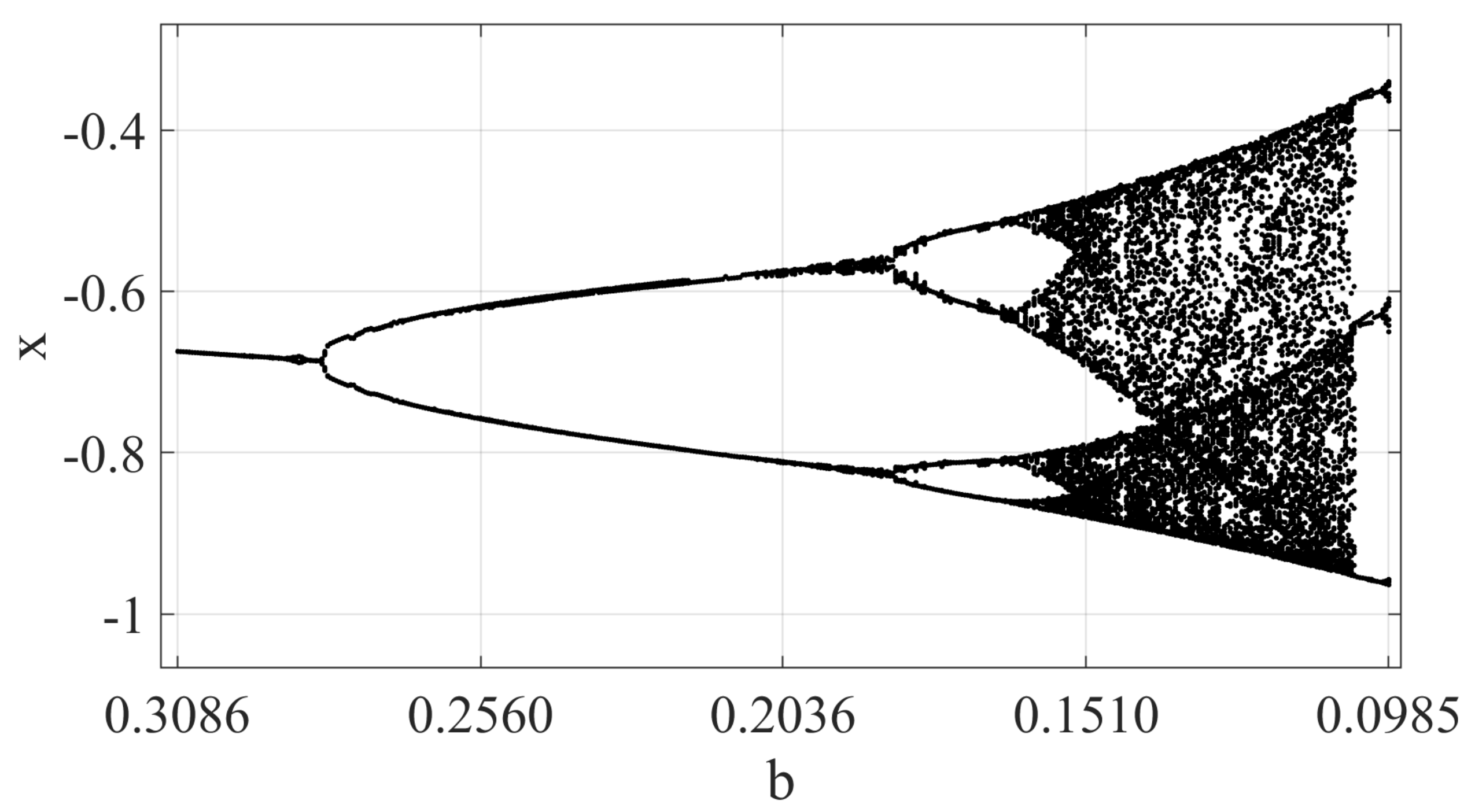}
			\caption{Numerical bifurcation diagram showing a period-doubling cascade with the change of parameter $b\in[0.0985,0.3086]$. The $x$-axis has been horizontally flipped to match the experimental diagram.}
			\label{rb_num}
		\end{subfigure}
        \hfill
        % \vspace{0.2cm}
		\begin{subfigure}{\columnwidth}
        \centering
    		\includegraphics[height=0.45\textwidth,width=0.6\textwidth]{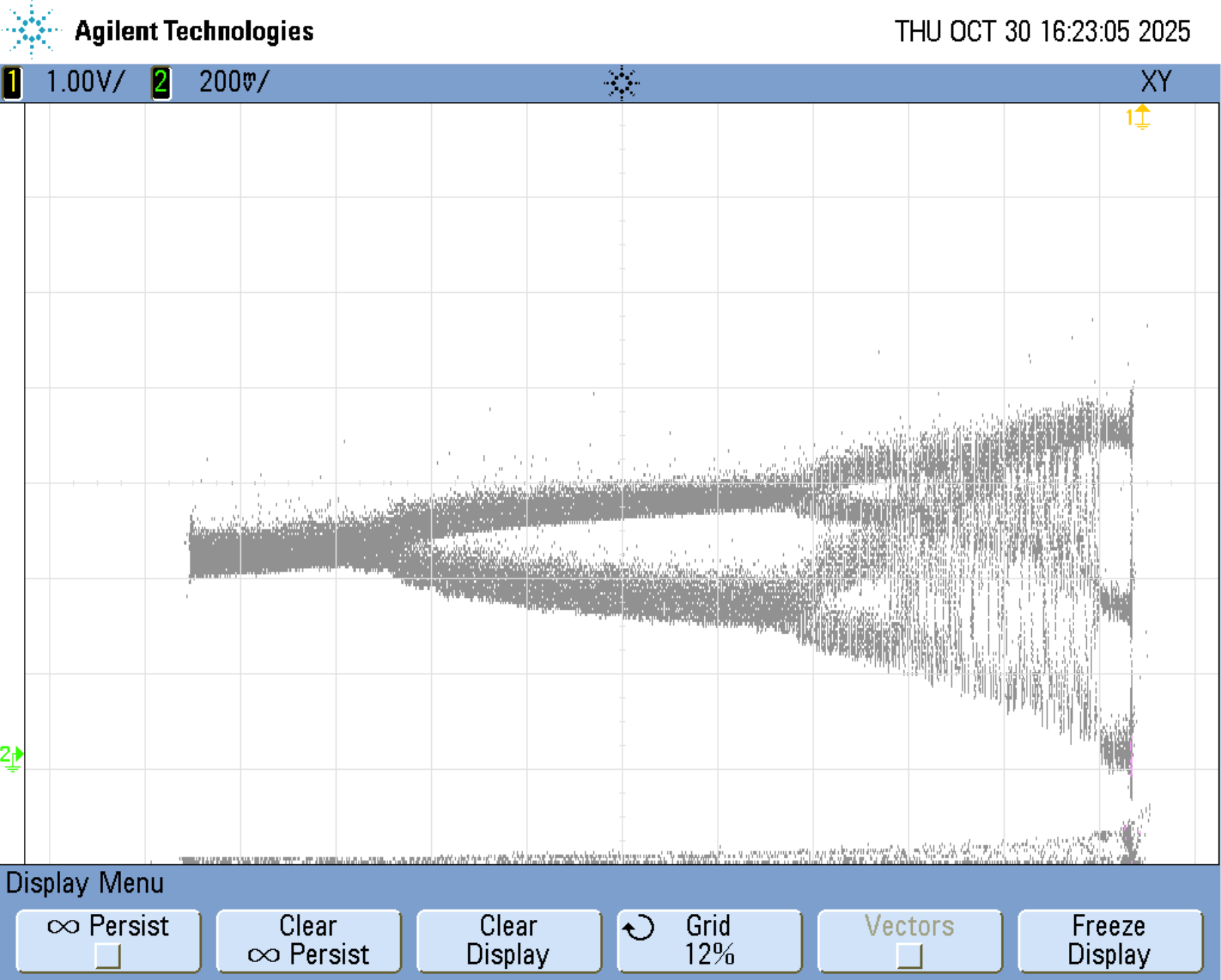}
		      \caption{Experimentally obtained bifurcation diagram using the \emph{persist} mode of the oscilloscope. The $x$-axis corresponds to voltage in the range $[-10~\text{V},-0.1~\text{V}]$ equivalent to $b$ being varied from $0.3086$ to $0.0031$, while, the $y$-axis is voltage corresponding to $x$ variable.}
		      \label{rb_exp}
		\end{subfigure}
 		\caption{ Numerical and experimental regular bifurcation diagrams for the rescaled R\"ossler system~(\ref{rosseqn2}). It is evident that the parameter ranges of (a) and (b) do not identically match; however, the qualitative behavior remains the same. Constant parameters: $a = 0.2$ and $c = 7.57$.}
		\label{rb}
	\end{figure}

We first obtain the regular bifurcation diagram numerically by varying $b$ from $0.0985$ to $0.3086$ (Fig.~\ref{rb_num}). For the experimentally obtained bifurcation diagram (Fig.~\ref{rb_exp}), $b$ is varied by the DAQ as a DC voltage ($V_{in}$) in the range $[-10~\text{V},-0.1~\text{V}]$  using the relation $b = - \tfrac{10}{324}\times V_{\rm in}$ (where the coefficient comes from the scaling between the input voltage and the parameter $b$, determined by the resistor values in the circuit). Comparing the two diagrams, we see that component tolerance and noise introduce quantitative differences in behavior; however, the period-doubling cascade remains intact.

\subsection{Spectral bifurcation diagram}
\begin{figure}[ht]
    \centering
    \includegraphics[width=0.7\linewidth]{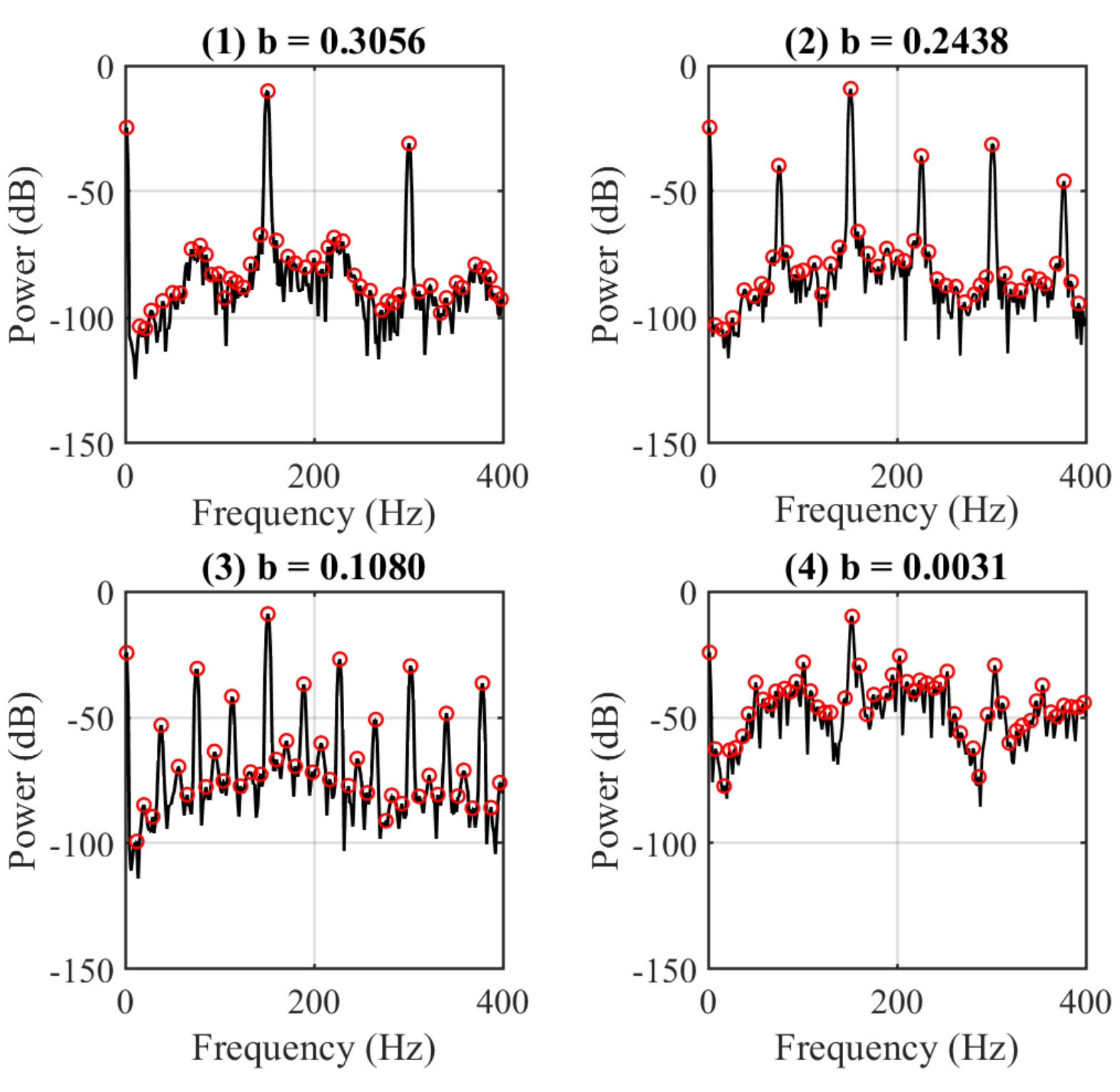}
    \caption{FFT spectra of system~\ref{rosseqn2} obtained from experimental data as $b$ is varied. (1) shows period-1, (2) shows period-2, (3) shows period-4, and (4) shows chaos. The peaks identified by our algorithm for plotting the SBD are marked with red circles. Constant parameters: $a = 0.2$ and $c = 7.57$.}
    \label{FFTs}
\end{figure}

Using DAQ, we extract time-series data as $b$ varies. This data is then used to generate FFT for different parameter values, as shown in Fig.~\ref{FFTs}. We see a clear halving of frequencies as the periodicity doubles.

\begin{figure}[ht]
	\centering
	\begin{subfigure}[]{\columnwidth}
		\centering
		\includegraphics[width=0.9\linewidth]{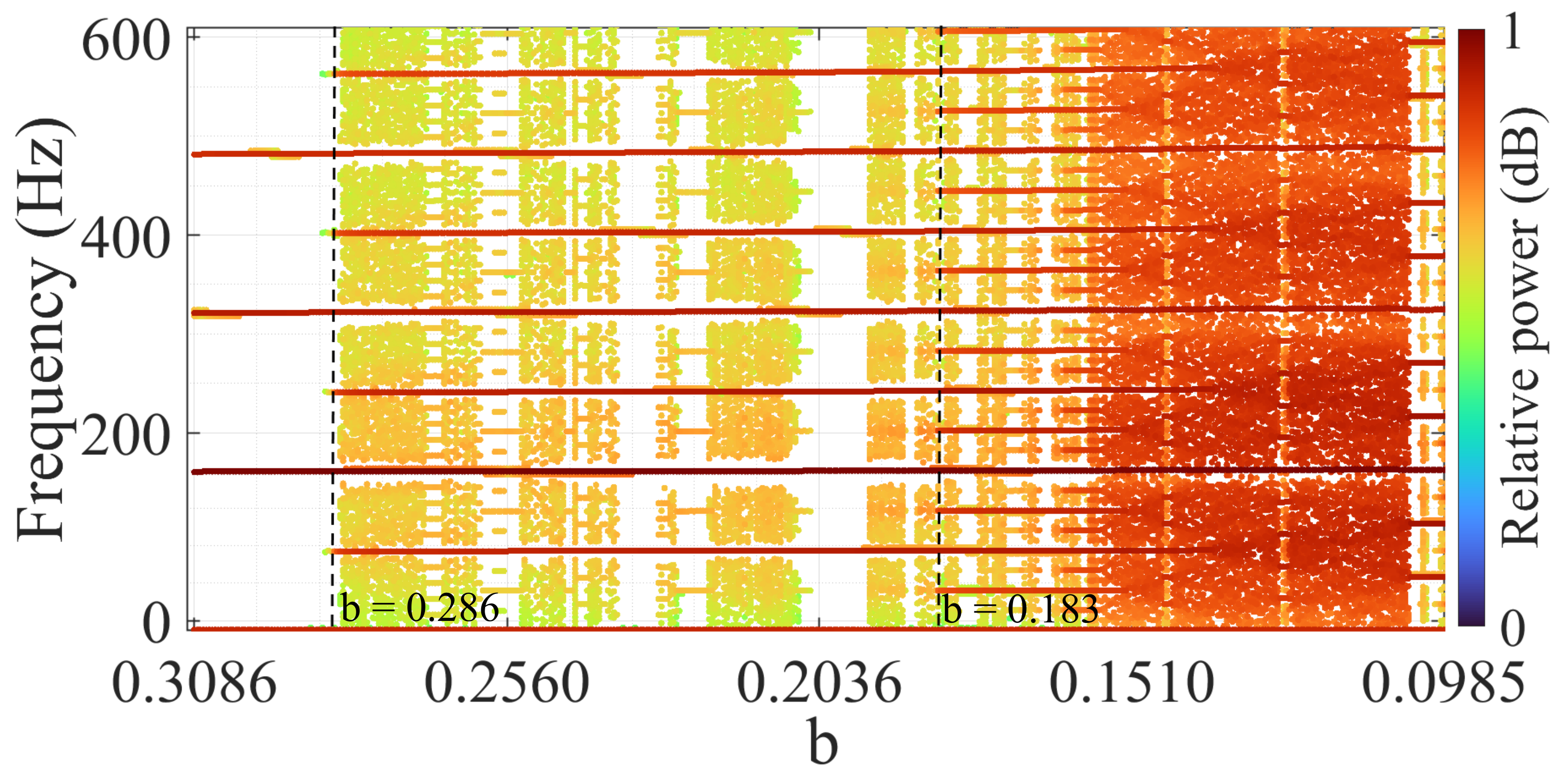}
		\caption{Numerically obtained SBD with $x$-axis flipped (to compare with experimental SBD) and the unbounded frequency axis cropped at $600$~Hz. }
		\label{rsbd_num}
	\end{subfigure}
    \hfill
    % \vspace{0.2cm}
	\begin{subfigure}[]{\columnwidth}
	\centering
		\includegraphics[width=0.9\linewidth]{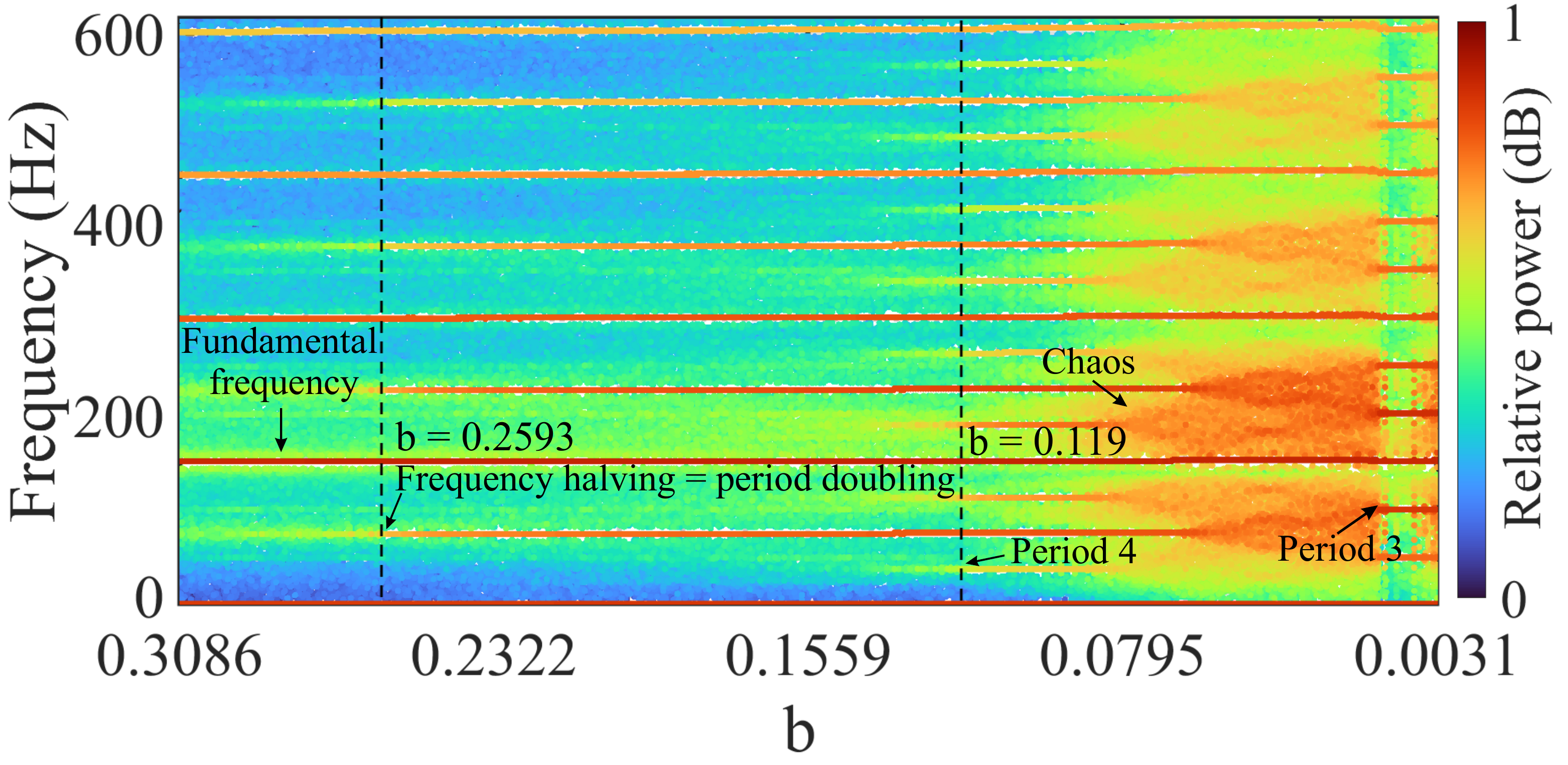}
		\caption{Experimentally obtained SBD. The voltage values $[-10~\text{V},-0.1~\text{V}]$ have been mapped to $b$ varying from 0.3086 to 0.0031.}
		\label{rsbd_exp}
		\end{subfigure}				
	\caption{Comparison of experimental and numerical SBDs of system~(\ref{rosseqn2}).}
	\label{Spectral Bifurcation Diagram}
\end{figure}

To study how this change unfolds, we filter the peak frequencies whose power exceeds a threshold of $-140~\text{dB}$~\cite{GU24}, as marked in Fig.~\ref{FFTs}. Plotting these peak frequencies as $b$ varies produces the SBD. The SBD generated from the experimental data is shown alongside numerically simulated SBD (Fig.~\ref{Spectral Bifurcation Diagram}). 

We note that, in most cases, the numerical parameter range differs from the experimental range due to tolerances, convenient capacitor choices, etc. Thus, we place greater emphasis on comparing the frequency content of the two SBDs than on exact parameter-value matching.

The primary rule to analyze an SBD is that we consider the lowest prominent frequency (we call it the lowest `rung', following~\cite{GU24}) at a given parameter value in an SBD to be the fundamental frequency and the higher frequencies at integer multiples of this frequency to be harmonics. We focus our attention mainly on the fundamental frequencies. 

   \begin{figure}[b]
	   \centering
      
        \includegraphics[width= \linewidth]{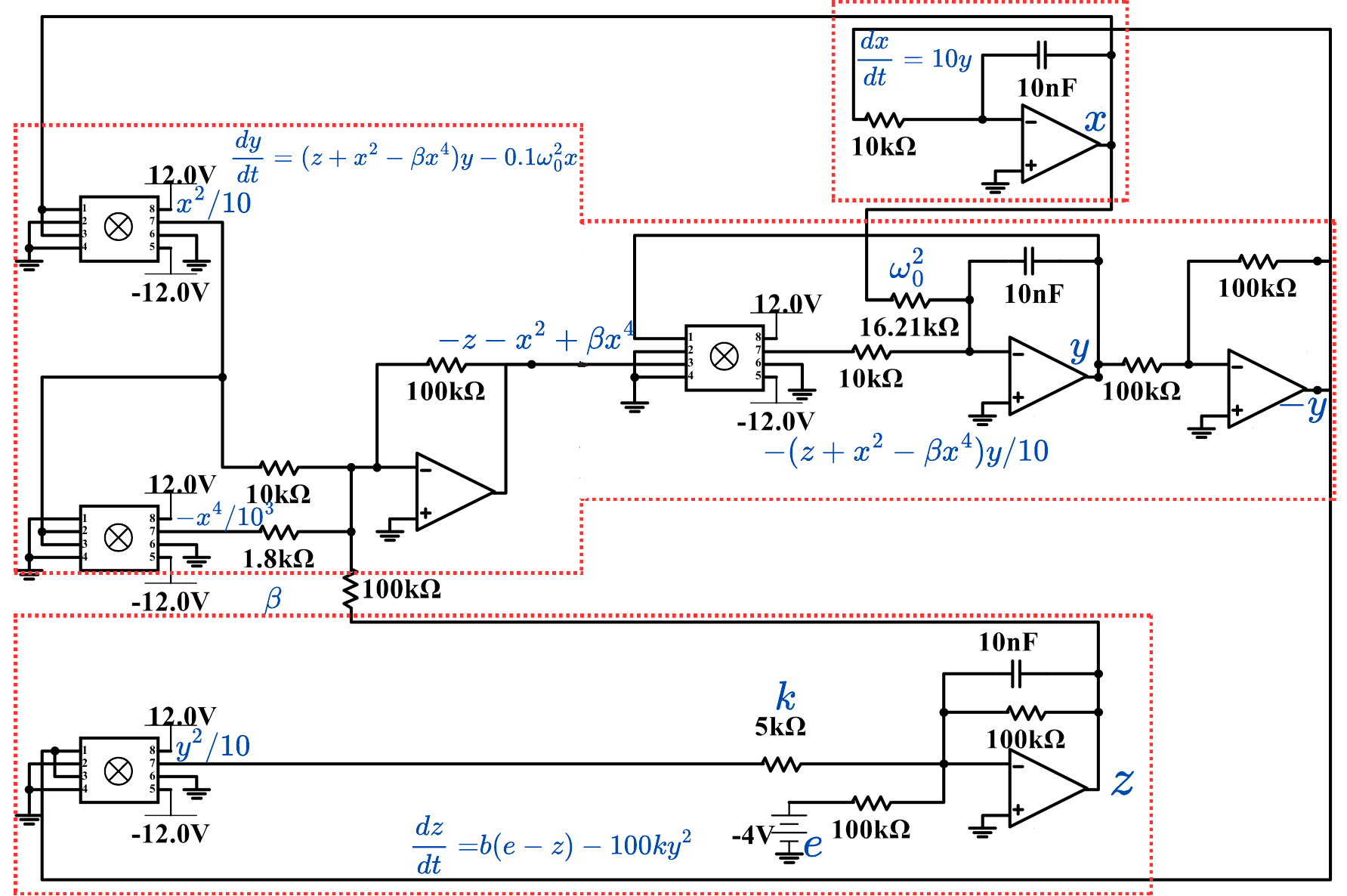}
	    \caption{Circuit implementation of the system (\ref{kuzeqn}). Control parameter $e$ is varied using DAQ output.}
	    \label{QP_circuit}
    \end{figure}

In Fig.~\ref{Spectral Bifurcation Diagram}, we see that at $b=0.3086$, the fundamental frequency is $\approx160~\text{Hz}$. As soon as the period doubles at $b\approx0.286$, a new frequency component $\approx80~\text{Hz}$ appears at half this value (Fig.~\ref{rsbd_num}). Comparing this with the experimental SBD in Fig.~\ref{rsbd_exp}, we see a slight mismatch where the halving seems to appear at $b\approx0.2593$. 

At b = 0.183, the frequency is again halved 
 to $\approx40~\text{Hz}$, indicating period-4.  In the experimental diagram, this is seen to happen at $b=0.119$. The halving continues, ultimately giving rise to a wide band of frequencies that represent chaos.

One may note that, unlike the numerical SBD, the experimental SBD is embedded in a sea of low-powered frequencies (blue) arising from noise coupled with the output. However, since it is not easy to decouple noise from a nonlinear response system without risking loss of spectral data, we did not use a stricter threshold when generating the experimental SBDs.

\section{Experimental SBD analysis of various phenomena} \label{exp_analysis}

We now demonstrate how this methodology could be used to analyze systems showing various dynamical behaviors, namely: two-frequency quasiperiodicity, torus length-doubling, and three-frequency quasiperiodicity.

	\subsection{Two-frequency quasiperiodicity} \label{quasisecn}
    
	Quasiperiodicity involves the interplay between two or more incommensurate frequencies. The FFT spectrum in such a case consists of frequencies in the form $mf_1\pm~nf_2\pm\cdots$~\cite{hilborn2000chaos,kaneko1984oscillation,glazier1988quasi,suarez2012stability}, where $f_i$ ($i = 1,2,\cdots$) denotes the incommensurate frequencies.

	   \begin{figure}[t]
	       \centering
	\fbox{\includegraphics[width=0.4\linewidth]{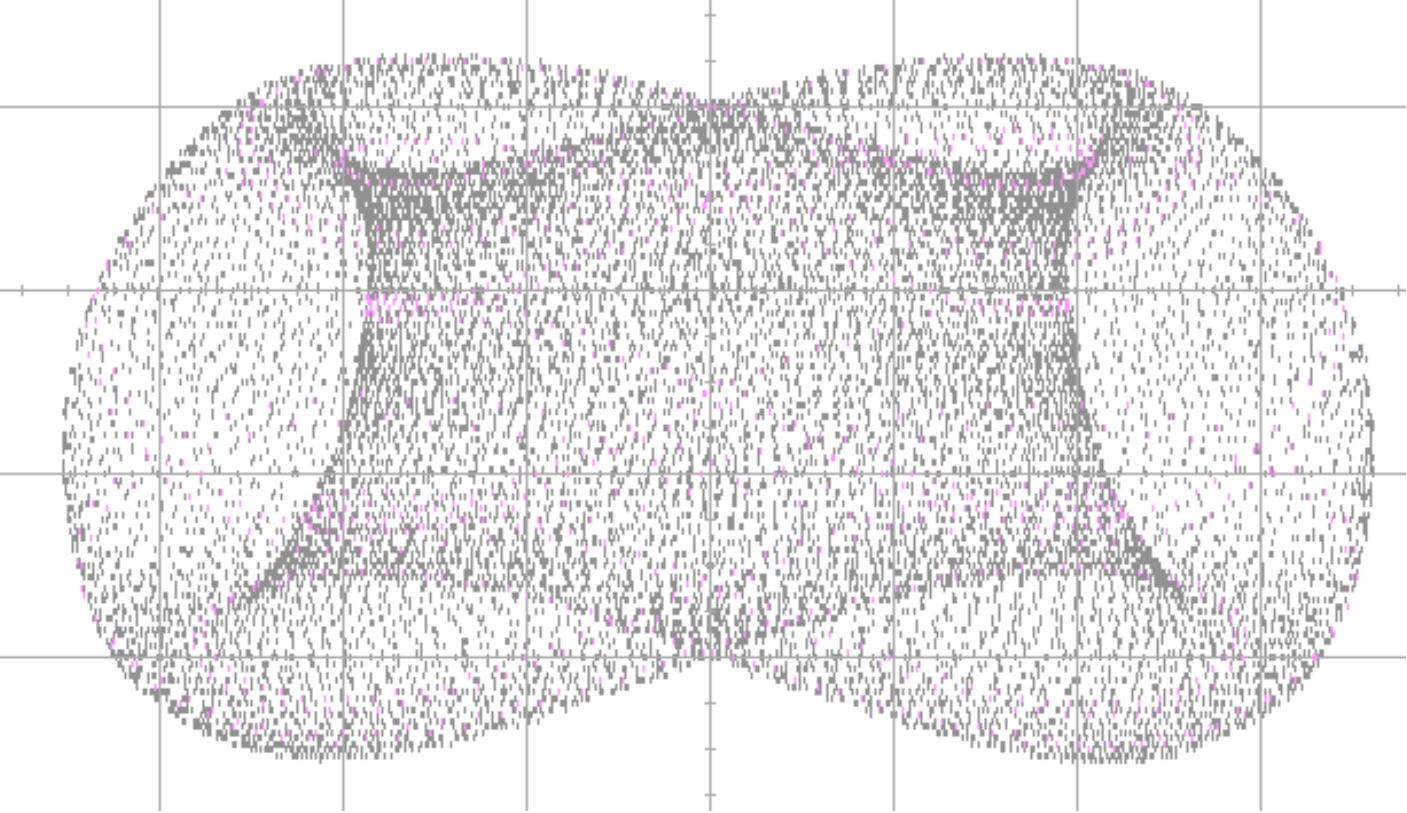}}
     		\caption{Experimentally obtained phase portrait (plotted $y$ vs. $z$) at $e = 5$.}
		\label{QP}
	\end{figure}

    We choose a 3-D oscillator system proposed by Kuznetsov et al.~\cite{KU13}. 
     The equations, after being scaled down by a factor of $10$ to avoid saturation, are given by (\ref{kuzeqn}):
	\begin{equation} \label{kuzeqn}
         \begin{array}{lll}
			\dot{x} &=& 10y\\
			\dot{y} &=& (\lambda + z + x^2 - \beta x^4)y -0.1 \omega_0^2 x\\
			\dot{z} &=& b(e-z) - 100ky^2
         \end{array}
	\end{equation}
    
    The system exhibits a secondary Hopf bifurcation, transitioning from periodic to quasiperiodic behavior, as the control parameter $e$ is varied over $2.4\le e \le 6$. 
  The fixed parameters are
	$\lambda = 0,~\beta =\tfrac{1}{18},~\omega_0 = 2.5\pi, ~b = 1,~ k = 0.02$.    
	The circuit diagram corresponding to the differential equation is shown in Fig.~\ref{QP_circuit}, and the experimental phase portrait showing a quasiperiodic orbit is presented in Fig.~\ref{QP}.

\subsubsection*{Bifurcation diagram}
\begin{figure}[ht]
		\centering
		\begin{subfigure}{\columnwidth}
			\centering
		\includegraphics[height=0.47\textwidth,width=0.77\linewidth]{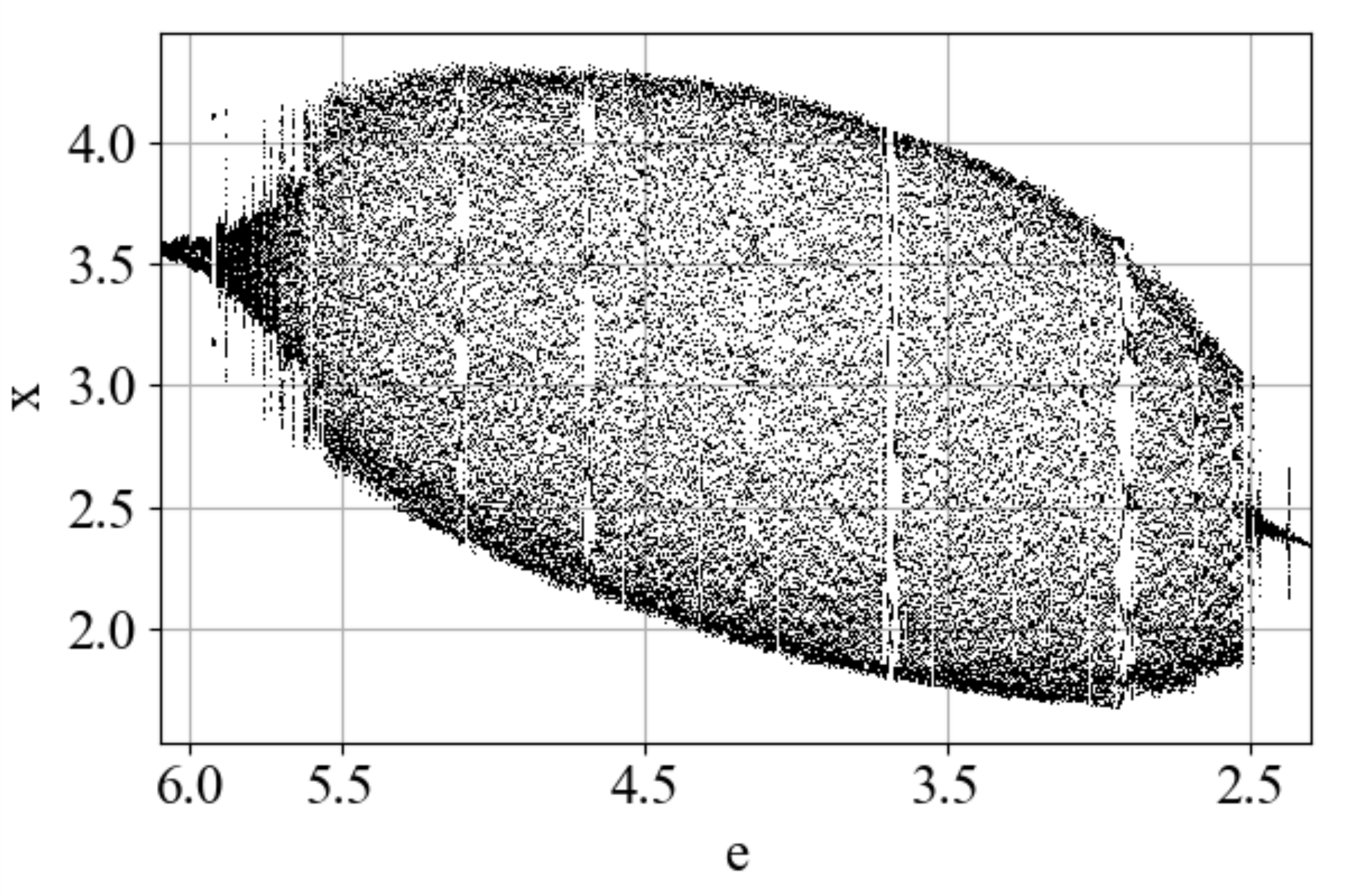}
			\caption{Numerical bifurcation diagram ($x$-axis flipped) $e\in[2.3,6.1]$.}
			\label{QP_bifur_num}
		\end{subfigure}
		\hfill
        % \vspace{0.2cm}
		\begin{subfigure}{\columnwidth}
			\centering
			\includegraphics[height=0.4\textwidth,width = 0.75\linewidth]{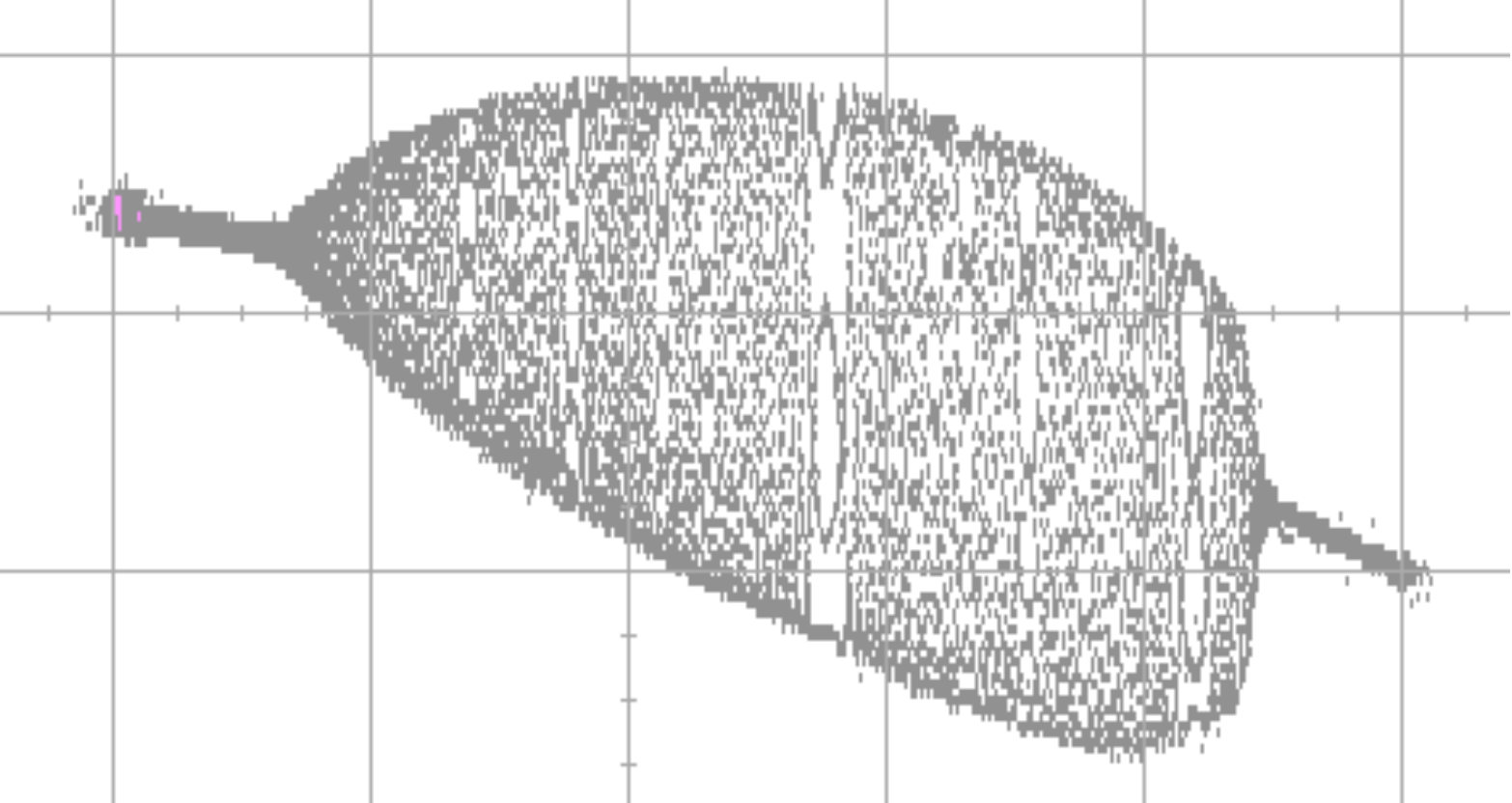}
			\caption{Experimental bifurcation diagram ($x$-axis is voltage V $\in$ $[-6.4~\text{V},-1.8~\text{V}]$ and $y$-axis is voltage corresponding to $x$).}
			\label{QP_bifur_exp}
		\end{subfigure}
		
		\caption{Qualitative comparison between the numerically obtained and the experimentally obtained bifurcation diagrams for system (\ref{kuzeqn}). Constant parameters: $\lambda = 0,~\beta =\tfrac{1}{18},~\omega_0 = 2.5\pi, ~b = 1,~ k = 0.02$.}
		\label{QP_Bifurcations}
	\end{figure}

As the control parameter $e\in[1.8,6.4]$ is varied, the numerical bifurcation diagram (Fig.~\ref{QP_bifur_num}) shows the creation and annihilation of a torus occurring in $e\approx6$ and $e\approx2.5$, respectively. Between these parameter values, the system exhibits quasiperiodic behavior and transitions to mode-locked periodic behavior over small ranges of $e$. 
    
In the experiment, the same parameter range is obtained for the voltage variation $[-1.8~\text{V},-6.4~\text{V}]$ as shown in Fig.~\ref{QP_bifur_exp}, where the DAQ voltage ($V_{in}$) and the parameter $e$ are related by $e = -V_{in}$.

	% Bifurcation diagram was here.
    
	\subsubsection*{Spectral bifurcation diagram}
    
\begin{figure}[t]
		\centering
		
		\begin{subfigure}{\columnwidth}
			\centering
			\includegraphics[width=0.9\linewidth]{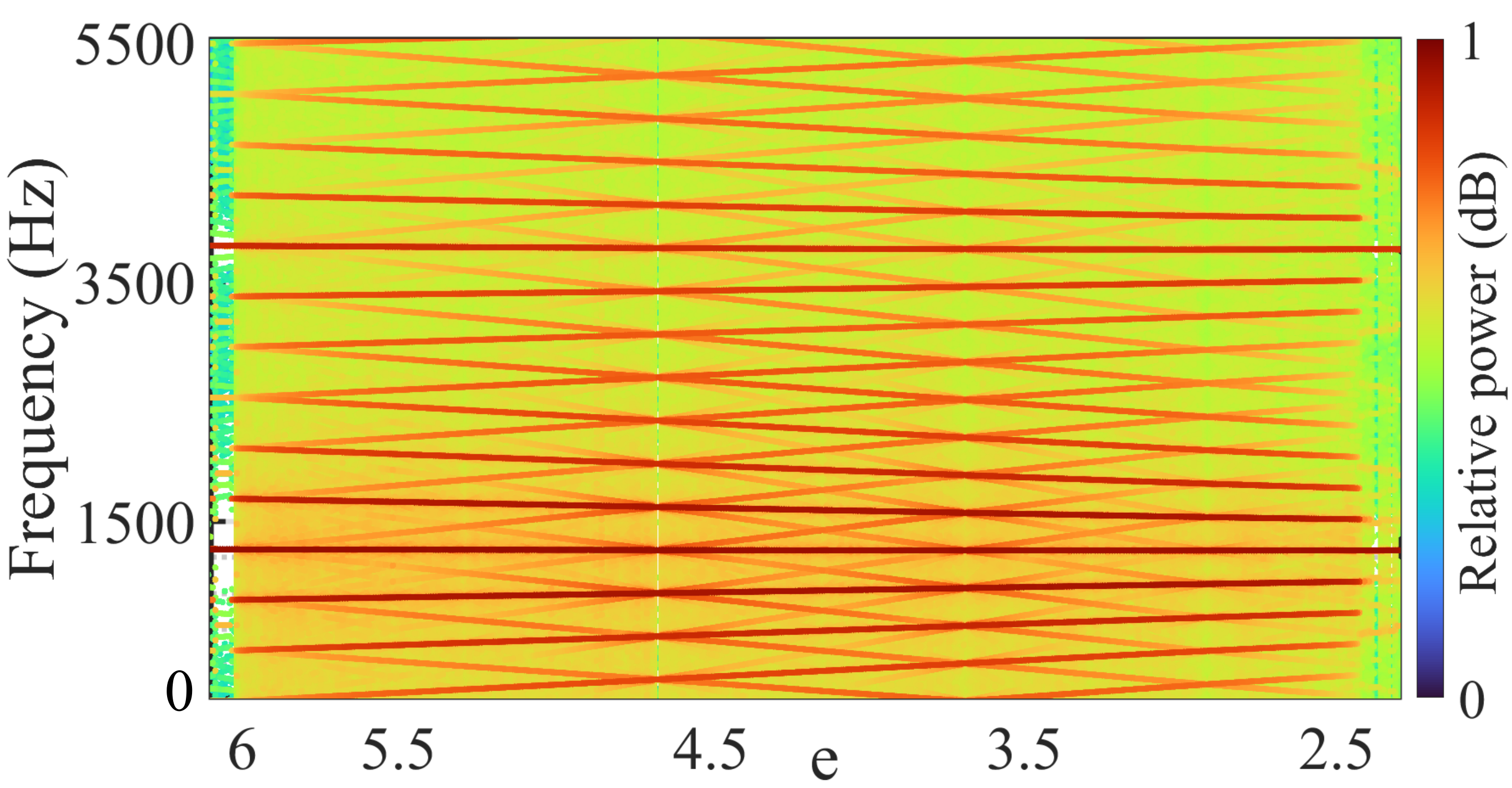}
			\caption{Numerical SBD for $e\in[2.3,6.1]$. }
			\label{qsbd_num}
		\end{subfigure}
        \hfill
		% \vspace{0.2cm}
		\begin{subfigure}{\columnwidth}
			\centering
			\includegraphics[width=0.9\linewidth]{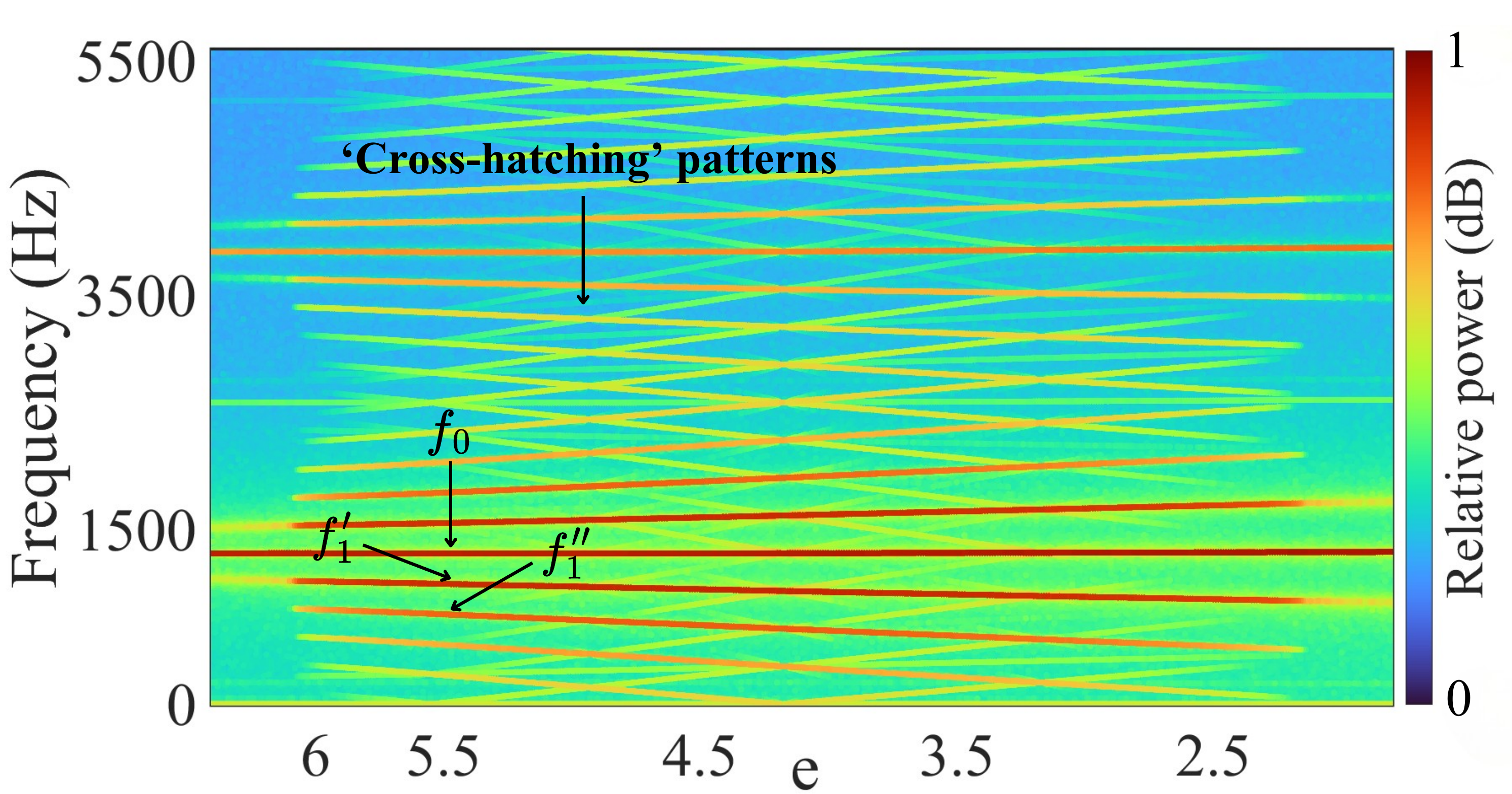}
			\caption{Experimental SBD for $e\in[1.8,6.4]$. $f_0=1238~\text{Hz,}~f_1'=1010~\text{Hz, and}~f_1''=782~\text{Hz}$.}
			\label{qsbd_exp}
		\end{subfigure}
		\hfill
		% \vspace{0.2cm}
		\begin{subfigure}{\columnwidth}
			\centering
			\includegraphics[width=0.9\linewidth]{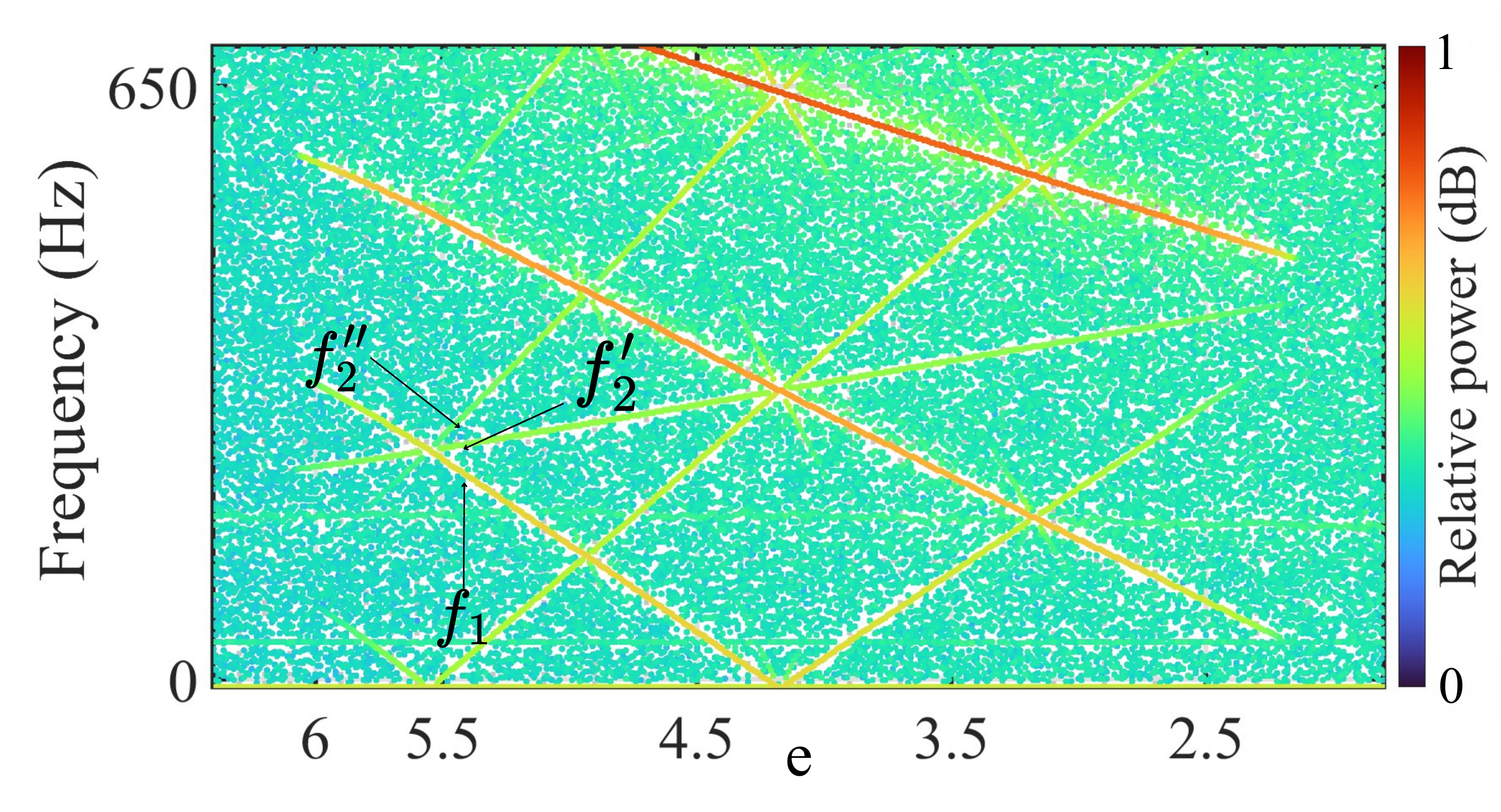}
			\caption{Vertically zoomed picture of experimental SBD. $e\in[1.8,6.4]$. $f_1=228~\text{Hz,}~f_2'=264~\text{Hz, and}~f_2''=299~\text{Hz}$. }
			\label{qpsbd_expzoomed}
		\end{subfigure}
		\caption{Comparison of numerical and experimental SBDs for quasiperiodic system (\ref{kuzeqn}).}
		\label{QP_SBD}
	\end{figure}

 	The numerical and experimental SBDs are plotted in Fig.~\ref{QP_SBD}.
 In both SBDs, there is only one dominant frequency $f_0\approx 1238~\text{Hz}$ in the periodic region, while a pattern of cross-hatching emerges in the quasiperiodic region immediately after the bifurcation of the torus at $e\approx6.0$, comprising two fundamental frequencies of the form $nf_1\pm~mf_2$~\cite{GU24,kaneko1984oscillation}. Due to the interplay between two frequencies, it is often difficult to identify the dominant frequency from the lowest rungs. In such cases, we apply a trick to extract the two frequencies as discussed in ~\cite{GU24}. 

We first fix a parameter value at which the harmonics are dense and prominent. We vertically scan for a set of dominant branches that appear equidistant from each other at this parameter value. Usually, these are the harmonics of $f_1$; consequently, the difference between these subsequent equidistant branches yields $f_1$. In Fig.~\ref{qsbd_exp}, we choose $e=5.4$ for this purpose, where we have marked three subsequent equidistant branches. The consecutive differences $(f_1''-f_1')$ and ($f_1'-f_0$) yield $f_1=228~\text{Hz}$.
    
Now, for the second frequency $f_2$, we need to zoom in on one of these $nf_1$ branches and study the fine structure. As shown in Fig.~\ref{qpsbd_expzoomed}, we zoomed in on the lowest branch, which is actually $f_1=228~\text{Hz}$. The immediate next branches labelled $f_2'$ and $f_2''$ are equidistant from each other. These are $f_1+f_2$ and $f_1+2f_2$. Hence, the differences between them yield $f_2\approx35~\text{Hz}$ at $e=5.4$. Using the SBD, it is also possible to identify the parameter ranges for mode-locking, where the cross-hatching branches meet.

	\subsection{Double covering bifurcation of a torus}
	When a torus undergoes a doubling bifurcation, it can result in two possible scenarios. In the first, the length of the loop doubles, a phenomenon called `torus length-doubling' or double-covering bifurcation. In the second, two separate loops form, and iterates flip between them, a phenomenon termed the loop-doubling bifurcation~\cite{banerjee2012local,kaneko1984oscillation,kamiyama2014classification,anishchenko2006winding,yu2007theoretical}. Here, we study the former type using the system (\ref{aneqn}) ,by Anishchenko et al.~\cite{AN05}:	    
    \begin{equation} \label{aneqn}
        \begin{split}
            \dot{x} =& \ m x + y - x\varphi - d x^3 \\
		\dot{y} =&\  -x \\
		\dot{z} =&\  \varphi \\
		\dot{\varphi} =&\ -\gamma \varphi + \gamma \Phi(x) - g z
        \end{split}
    \end{equation}
	where 
    \begin{eqnarray*}
\Phi(x) &=& x(x-0.7), \quad \text{if }\; x>0.7 \\
        &=& 0, \quad \quad \quad \quad \quad\text{otherwise.}
\end{eqnarray*}
	
\begin{figure}[ht]
		\centering
		\includegraphics[width = 0.7\linewidth]{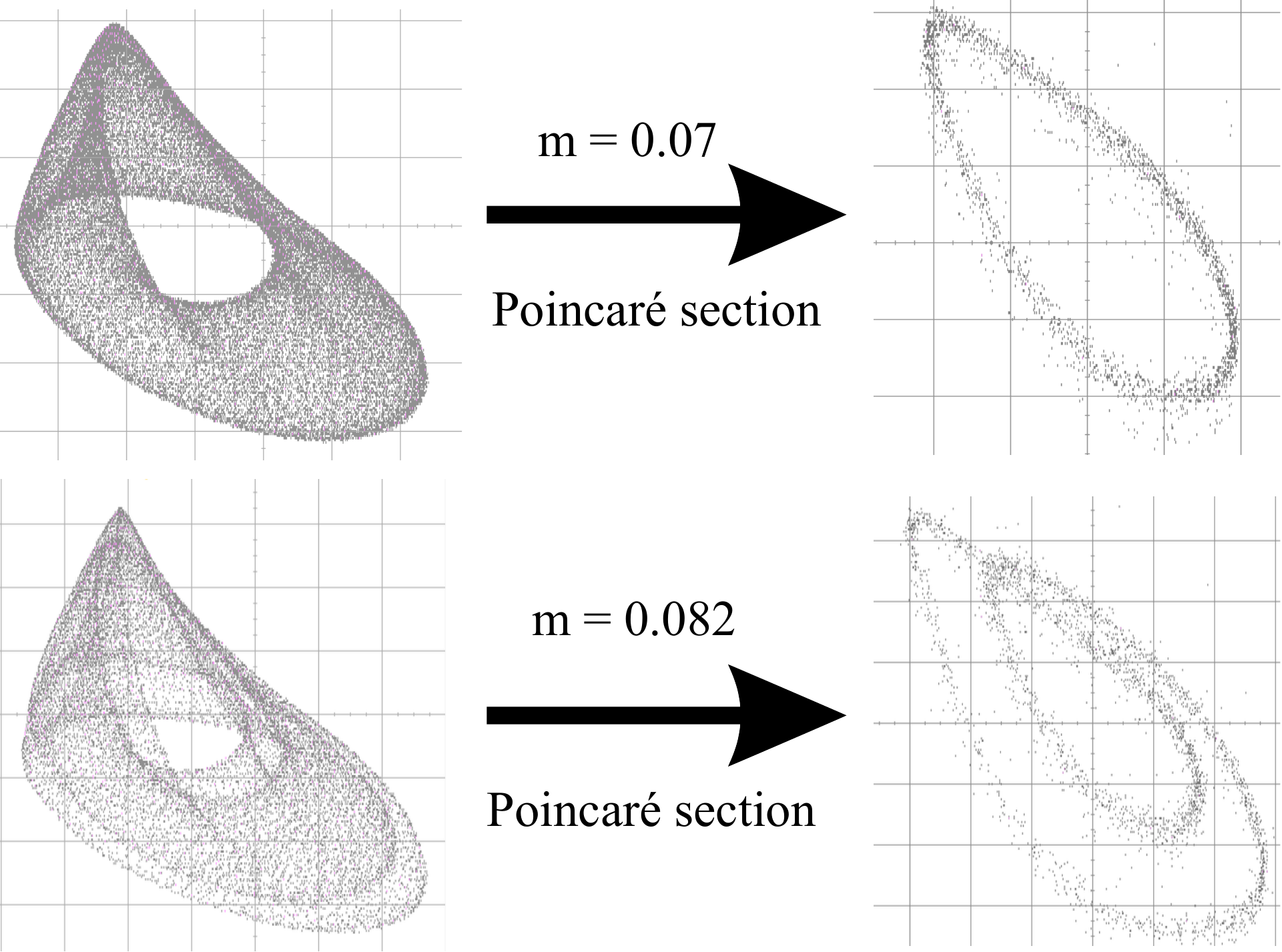}
		\caption{Experimental phase portraits ($y$ vs $z$) and their Poincar\'e sections showing the length-doubling of the loop as $m$ is varied in (\ref{aneqn}).}
		\label{ld1}
	\end{figure}

\begin{figure}[ht]
		\centering
		\includegraphics[width = 0.9\linewidth]{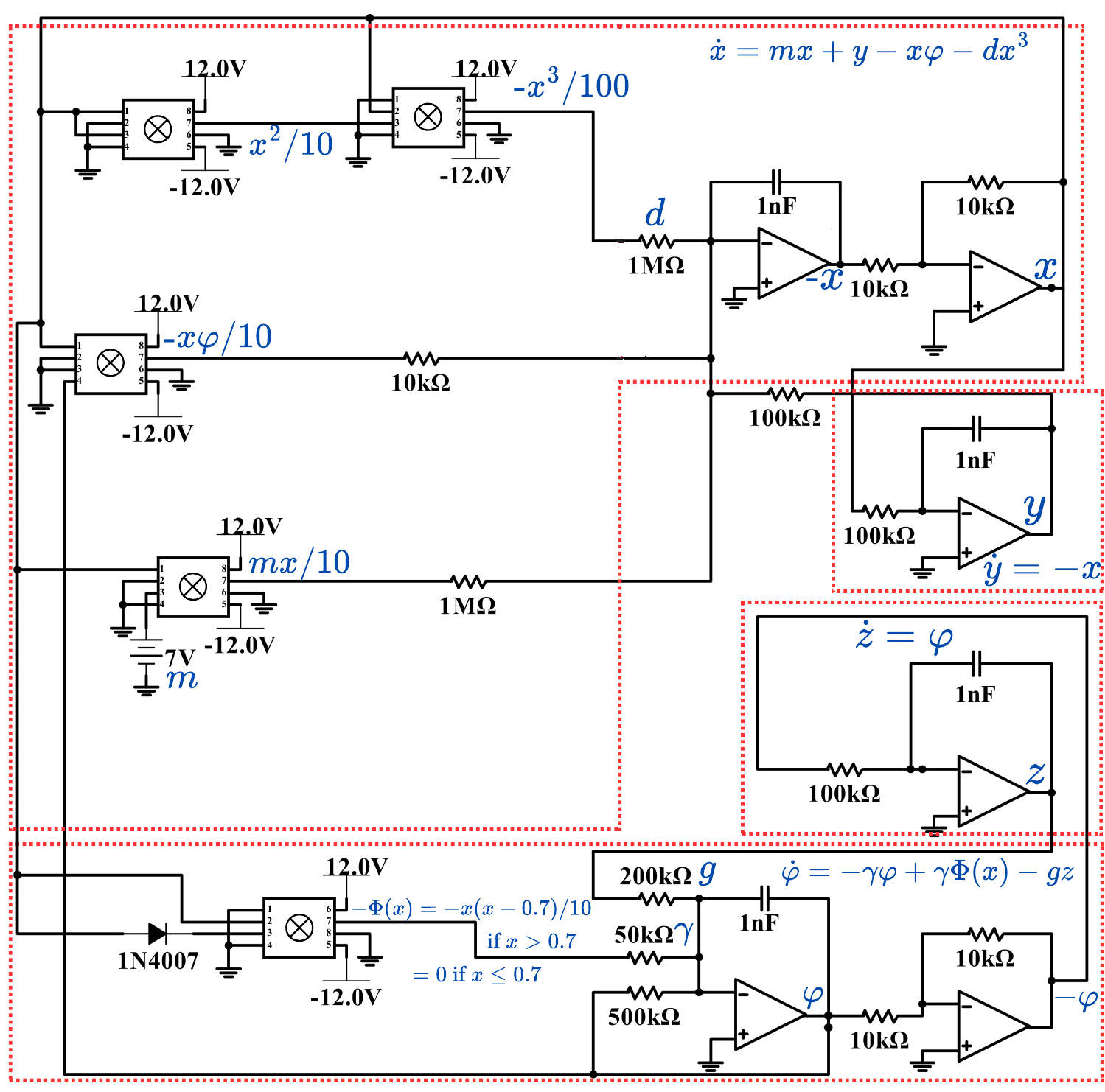}
		\caption{Circuit corresponding to system (\ref{aneqn}). Control parameter $m$ is varied using the DAQ output.}
		\label{ld_circuit}
	\end{figure}

The fixed parameters are
	$d = 0.001$, $g = 0.5$, $\gamma = 0.2$. The control parameter is $m$, varied in the range  
    $m\in[0.065,0.096]$, which is equivalent to the DAQ voltage variation $[6.5~\text{V},9.6~\text{V}]$. 
    The circuit corresponding to the system is shown in Fig.~\ref{ld_circuit}, where we use the linear approximation of an ideal diode to implement $\Phi(x)$~\cite{MA20}. The phase portraits and their Poincar\'e cross-sections are shown in Fig.~\ref{ld1}, where one can easily identify the length-doubling.

	\subsubsection*{Regular bifurcation diagram}
\begin{figure}[ht]
		\centering
		
		% --- Subfigure 1 ---
		\begin{subfigure}{\columnwidth}
			\centering
			\includegraphics[height=0.5\textwidth,width=0.8\textwidth]{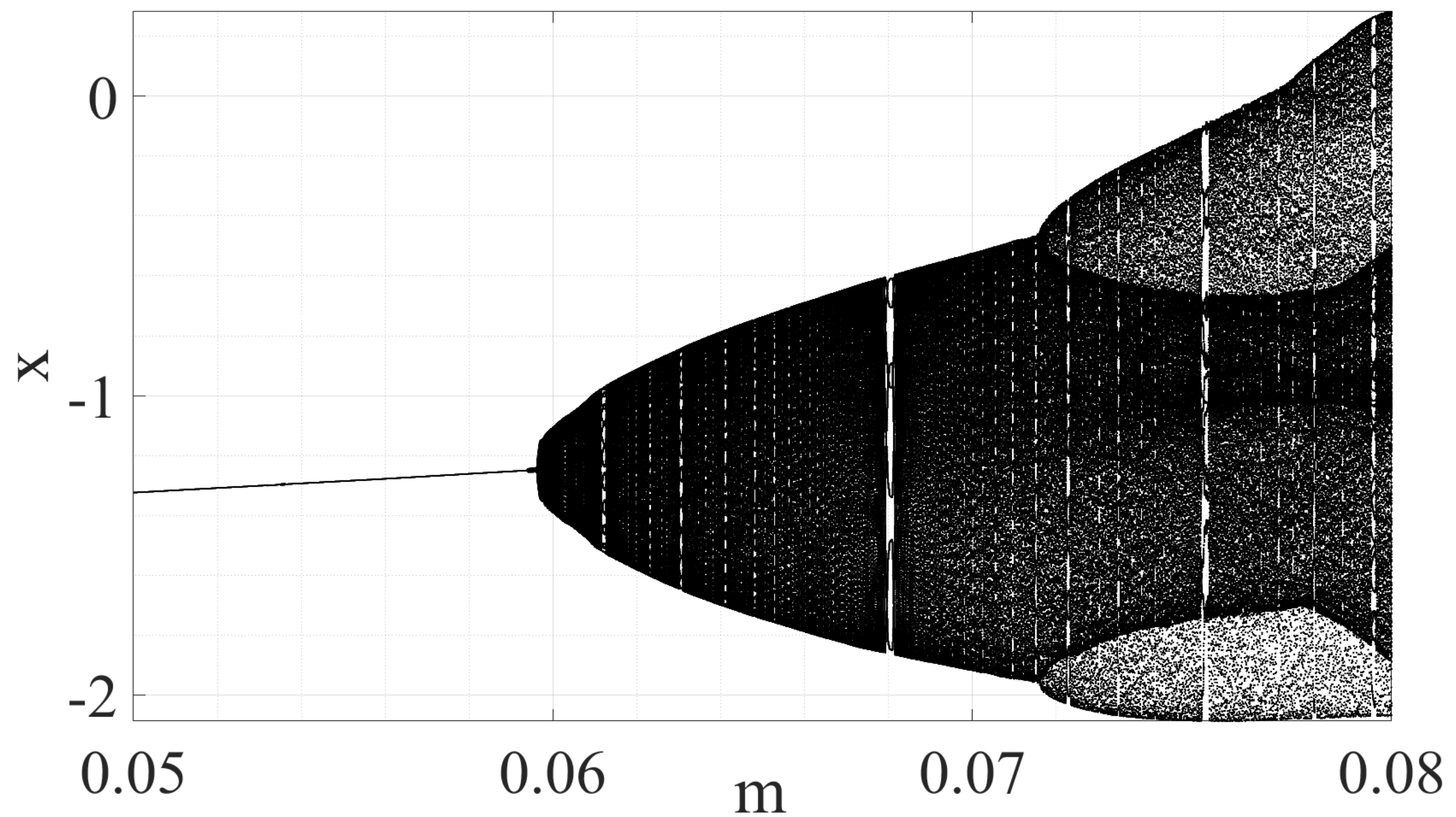}
			\caption{Numerical bifurcation diagram for $m\in[0.05,0.08]$.}
			\label{ld_num}
		\end{subfigure}
        \hfill
		% \vspace{0.2cm}
		% --- Subfigure 2 ---
		\begin{subfigure}{\columnwidth}
			\centering
			\includegraphics[height=0.5\textwidth,width= 0.8 \textwidth]{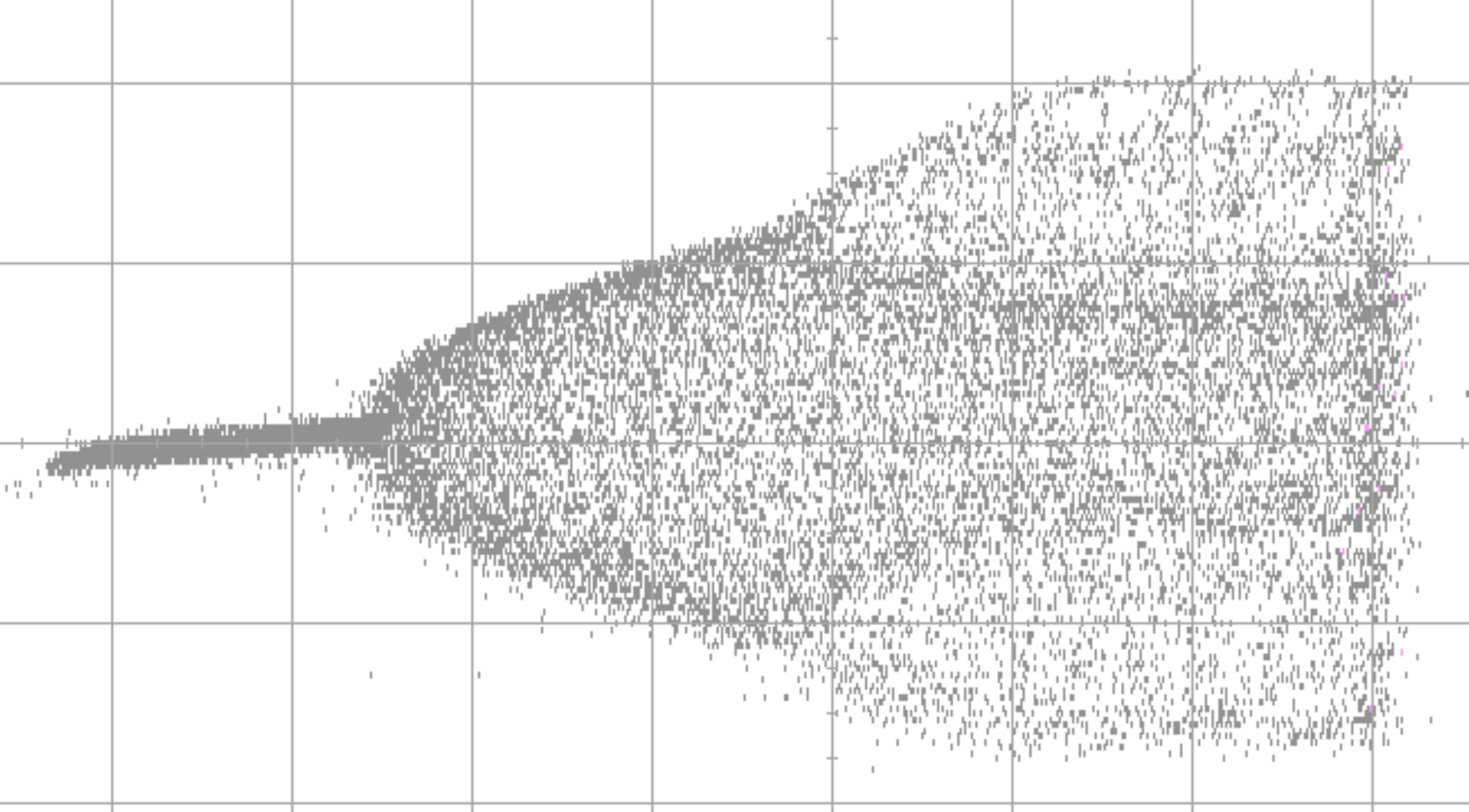}
			\caption{Experimental bifurcation diagram ($x$-axis is voltage V $\in$ $[6.5~\text{V},9.6~\text{V}]$) and $y$-axis is voltage corresponding to $x$.}
			\label{ld_exp}
		\end{subfigure}
		
		\caption{Comparing the numerically obtained bifurcation diagram with the one obtained experimentally for system (\ref{aneqn}). Constant parameters: $d = 0.001 ,~g = 0.5,~\gamma = 0.2$. }
		\label{ld_bifurcations}
	\end{figure}

	In Fig.~\ref{ld_bifurcations}, we see that the experimentally obtained bifurcation diagram agrees quite well with the numerically obtained one as $m$ is varied over $[0.065,0.096]$. Although the parameter regions do not exactly match due to the linear approximation of a nonlinear diode, the qualitative similarity allows us to extend the discussion to the SBDs (Fig.~\ref{LD_SBD}). 
	
	\subsubsection*{Spectral bifurcation diagrams}
\begin{figure}[tb]
		\centering	
		\begin{subfigure}{\columnwidth}
			\centering
			\includegraphics[width=0.9\textwidth]{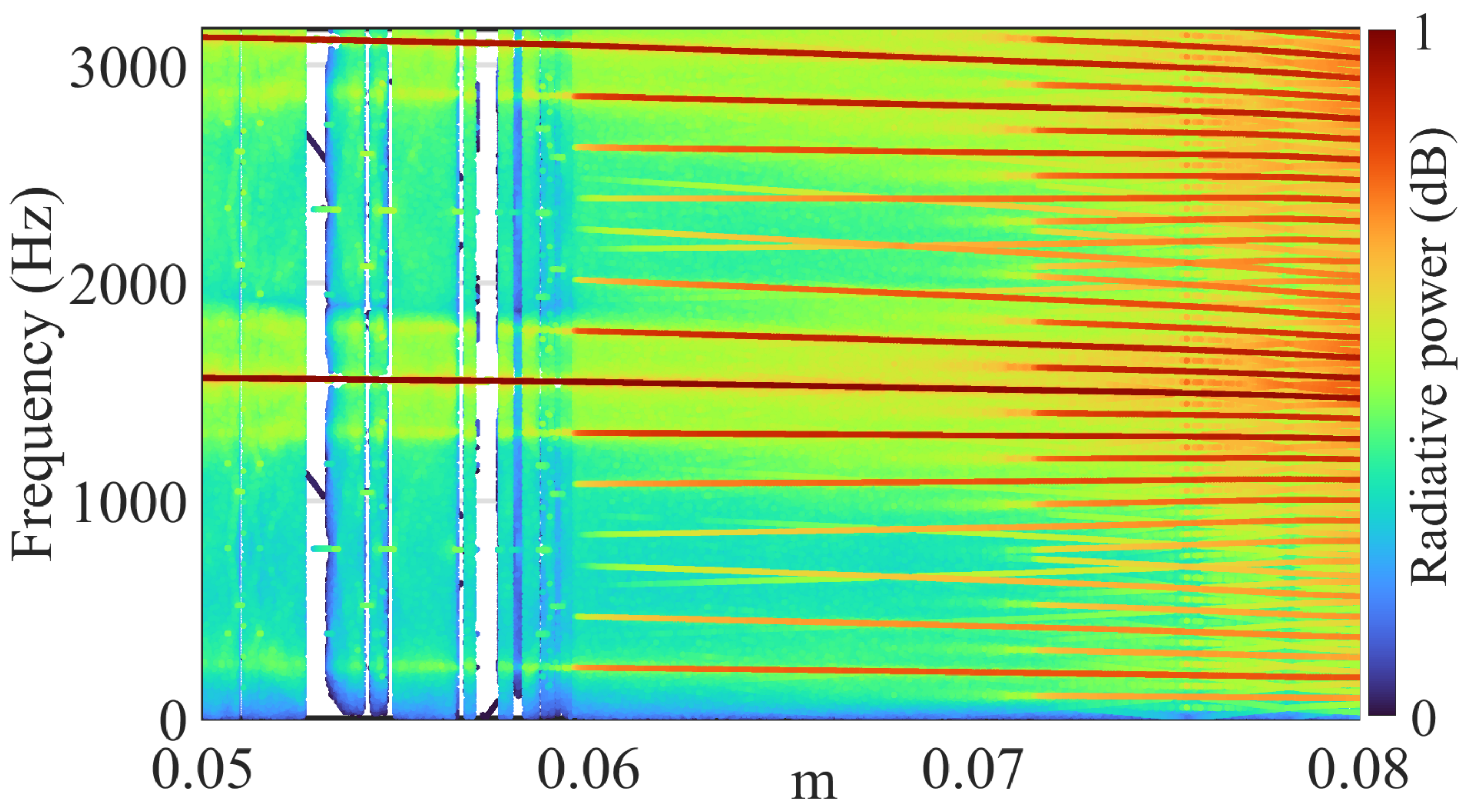}
			\caption{Numerical SBD for $m\in[0.05,0.08]$}
			\label{ldsbd_num}
		\end{subfigure}
	    % \vspace{0.2cm}
		% Subfigure 2 
		\begin{subfigure}{\columnwidth}
			\centering
			\includegraphics[width=0.9\textwidth]{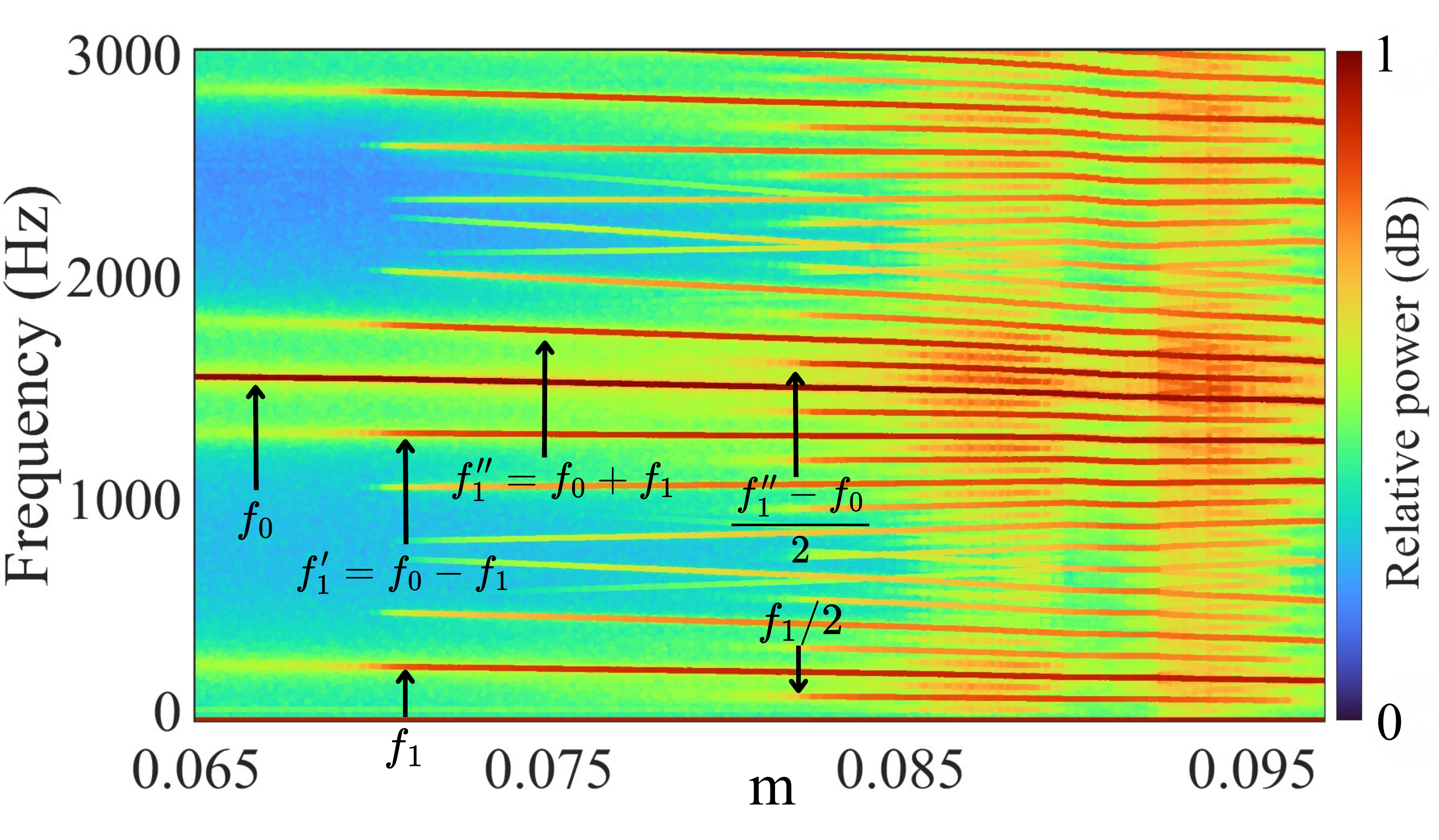}
			\caption{Experimental SBD for $m\in[0.065,0.096]$. Here, $f_0$ marks the periodic branch, while $ f_1$ marks one of the two quasiperiodic branches, and $f_1/2$ marks the frequency-halving following the torus-doubling. The rest of the spectrum is made of addition and subtraction of harmonics as shown.}
			\label{ldsbd_exp}
		\end{subfigure}
        % \vspace{2cm}
        % Subfigure 3 
        \begin{subfigure}{\columnwidth}
        \centering
		\includegraphics[width = 0.9\textwidth]{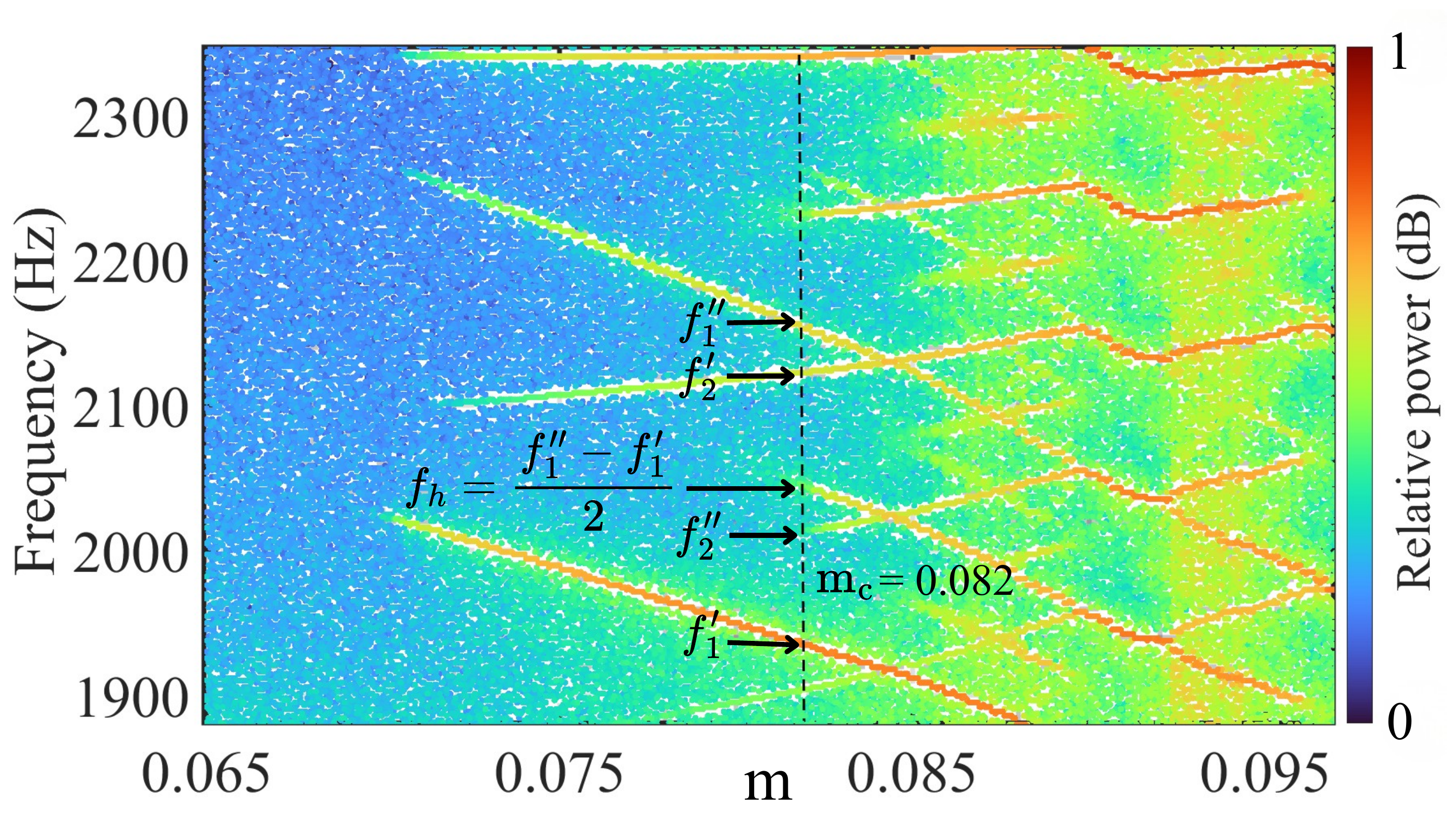}
		\caption{Fig.~\ref{ldsbd_exp} vertically zoomed. $f_2$, the other quasiperiodic frequency, is obtained from this zoomed image as $f_2=f_1''-f_2'$. $f_1'$ and $f_1''$ are subsequent harmonics of $f_1$ whose difference yields $f_1$, while $f_h=\tfrac{f_1}{2}$ is the halved frequency. }
		\label{ldsbd_expzoom}
        \end{subfigure}
        % \vspace{-2.3cm}
		\caption{Comparison of numerical and experimental SBDs for torus-doubling bifurcation seen in system (\ref{aneqn}).}
		\label{LD_SBD}
	\end{figure}

	In both SBDs (Figs.~\ref{ldsbd_num} and~\ref{ldsbd_exp}), we see a similar `cross-hatching' pattern in the quasiperiodic region, which, with further parameter changes, incorporates an intermediate frequency branch. 
    In the experimental SBD (Fig.~\ref{ldsbd_exp}), the periodic region $m<0.065$ comprises the frequency $f_0= 1545~\text{Hz}$. The doubling bifurcation is seen to occur at $m_c\approx0.082$ (shown in Fig.~\ref{ldsbd_expzoom}).  Just before the bifurcation ($m<m_c$), one can easily identify $f_1=218~\text{Hz}$. At $m=m_c$, we see that this frequency is halved to $\tfrac{f_1}{2}=109~\text{Hz}$, denoting a doubling.
	
    The other frequency, $f_2$, and its fate after bifurcation require a study of the fine structure of the SBD as shown in Fig.~\ref{ldsbd_expzoom}. With reference to the plot, the difference $f_1''-f_2'$ yields $f_2=31 ~\text{Hz}$ before the doubling bifurcation. After the bifurcation, we may obtain $f_2$ as the difference $f_h-f_2''= 31~\text{Hz}$, which turns out to be the same.
    This experimentally confirms that in a torus length-doubling bifurcation, one frequency is halved while the other remains unchanged, as reported in Guha and Banerjee~\cite{GU24}.
    
\subsection{Three-frequency quasiperiodicity}
\begin{figure}[tbh]
            \centering
            \includegraphics[width = \columnwidth]{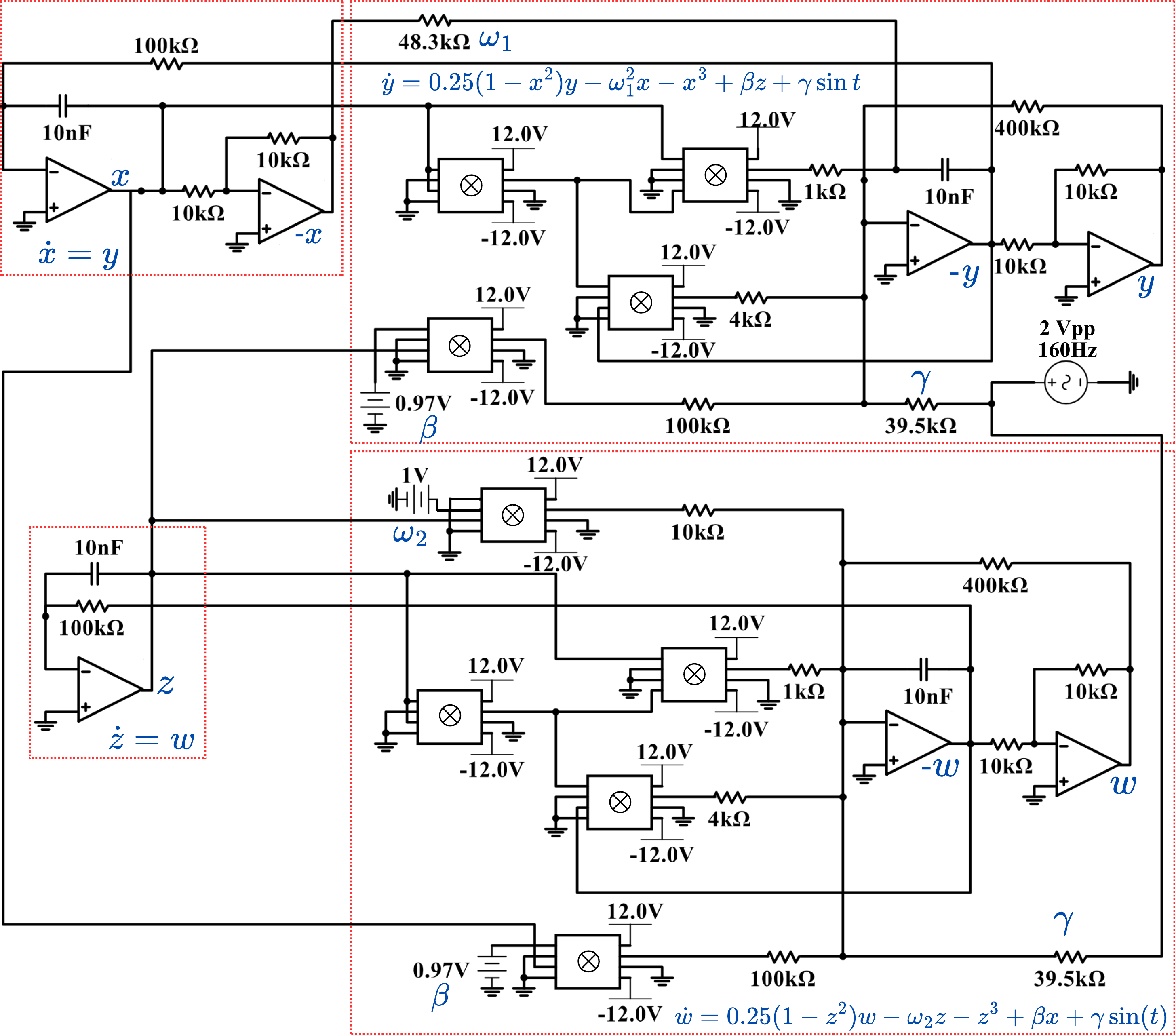}
            \caption{Circuit diagram for driven coupled Van der Pol oscillator (\ref{batteqn}). Control parameter $\omega_2$ is varied using the DAQ output. AC voltage is supplied by the MFG-2260MFA multi-channel function generator.}
            \label{tfq_CVP}
    \end{figure}

	If the torus undergoes a third consecutive Hopf bifurcation, a third frequency emerges, and the orbit lies on the surface of a 3-dimensional torus~\cite{giaouris2011complex}. To study this, we consider the driven coupled Van der Pol oscillator mentioned in~\cite{BA88}. The ODEs are given in (\ref{batteqn}), and the circuit used to simulate them is shown in Fig.~\ref{tfq_CVP}.
    
	\begin{equation} \label{batteqn}
		\begin{array}{lll}
			\dot{x} &=& y \\
			\dot{y} &=& 0.25(1 - x^2)y - \omega_1^2 x - x^3 + \beta z + \gamma \sin t \\
			\dot{z} &=& w \\
			\dot{w} &=& 0.25(1 - z^2)w - \omega_2 z - z^3 + \beta x + \gamma \sin t
		\end{array}
	\end{equation}
	
	\begin{figure}[b]
		\centering
		\includegraphics[width = \columnwidth]{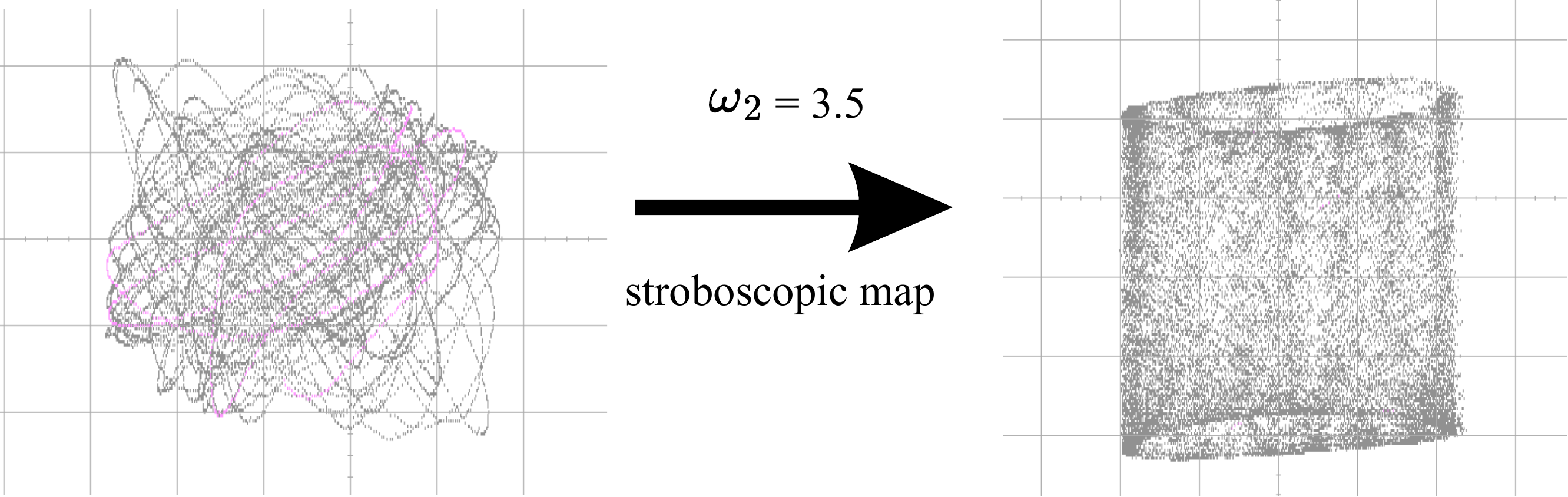}
		\caption{The orbit in phase space and the Poincar\'e section at $\omega_2 = 3.5$, as observed in the oscilloscope.}
		\label{tfq_strob}
	\end{figure}

    \subsubsection*{Regular bifurcation diagram}
\begin{figure}[ht]
		\centering
		\begin{subfigure}{\columnwidth}
			\centering
			\includegraphics[width=0.8\textwidth]{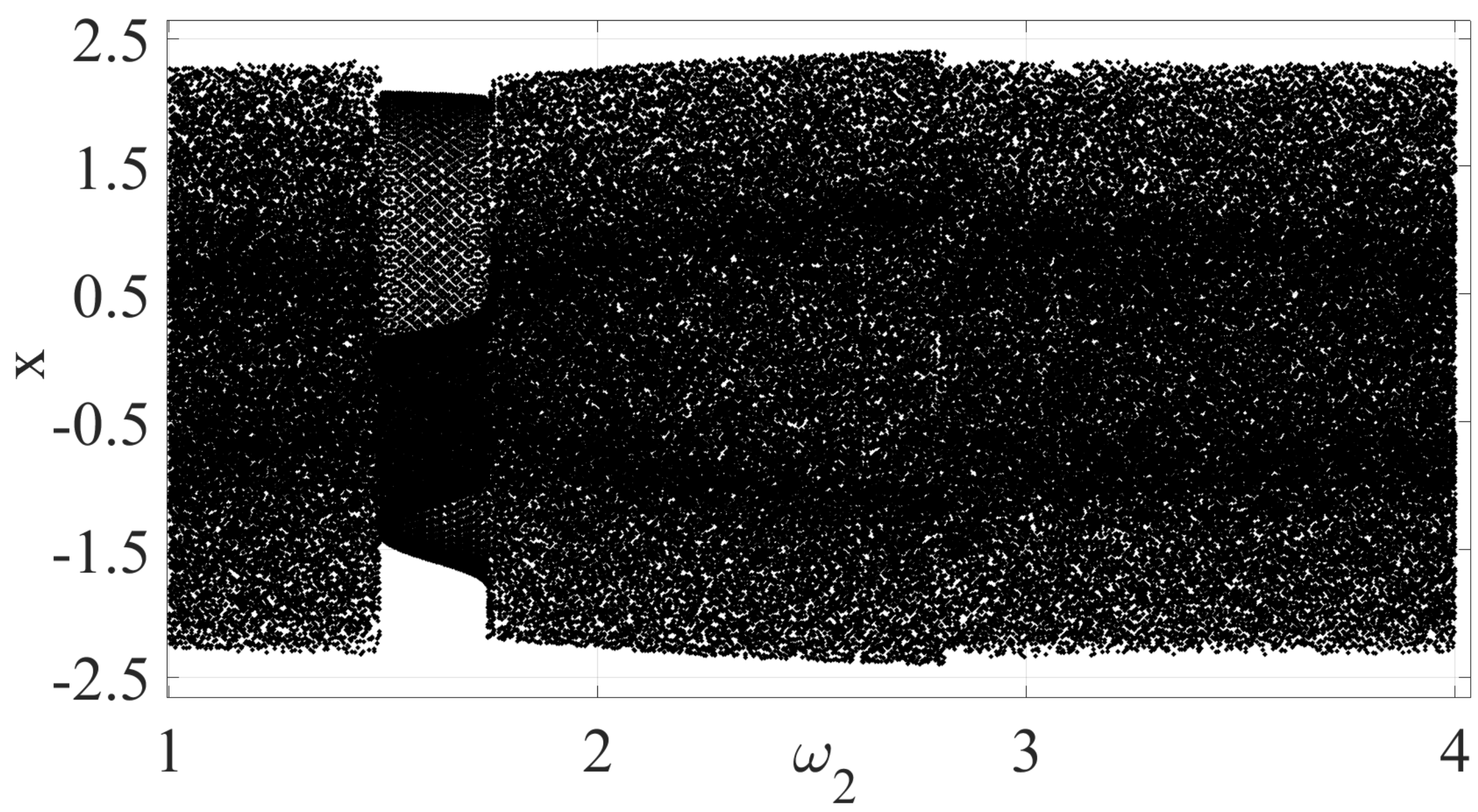}
			\caption{Numerical bifurcation diagram for $\omega_2\in[1,4]$.}
			\label{tfqb_num}
		\end{subfigure}
		% \vspace{0.2cm}
		\begin{subfigure}{\columnwidth}
			\centering
			\includegraphics[height=0.4\textwidth,width=0.9\textwidth]{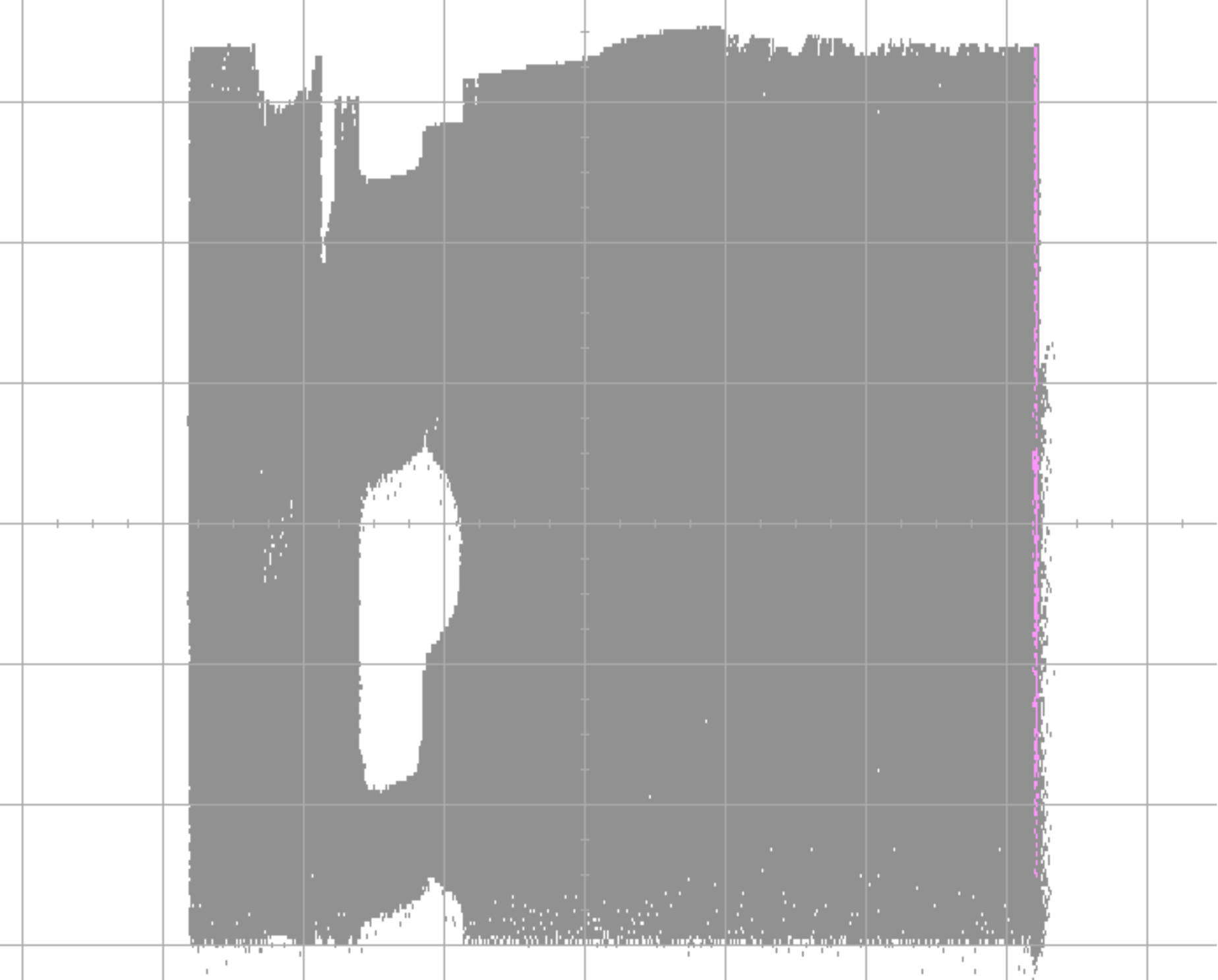}
			\caption{Experimental bifurcation diagram ($x$-axis is voltage V $\in [1~\text{V}, 4~\text{V}]$) and $y$-axis is voltage corresponding to $x$. }
			\label{tfqb_exp}
		\end{subfigure}
		\caption{Comparison of numerical and experimental bifurcation diagrams of system (\ref{batteqn}). Constant parameters: $\omega_1= 2.07,\beta = 0.097,\gamma = 2.53$. }
		\label{tfqb}
	\end{figure}

	For creating a bifurcation diagram (Fig.~\ref{tfqb}), $\omega_2$ is varied in $[1,4] $ (DAQ voltage varied in $[1~\text{V}, 4~\text{V}]$) while the other parameters are kept constant at $\omega_1= 2.07,\beta = 0.097,\gamma = 2.53$. Using a stroboscopic sampling in synchronism with the sinusoidal forcing, we can confirm the presence of 3-Torus as shown in Fig.~\ref{tfq_strob}.
	
	\subsubsection*{Spectral bifurcation diagram}
\begin{figure}[ht]
		\centering	
		\begin{subfigure}{\columnwidth}
			\centering
			\includegraphics[width=0.9\textwidth]{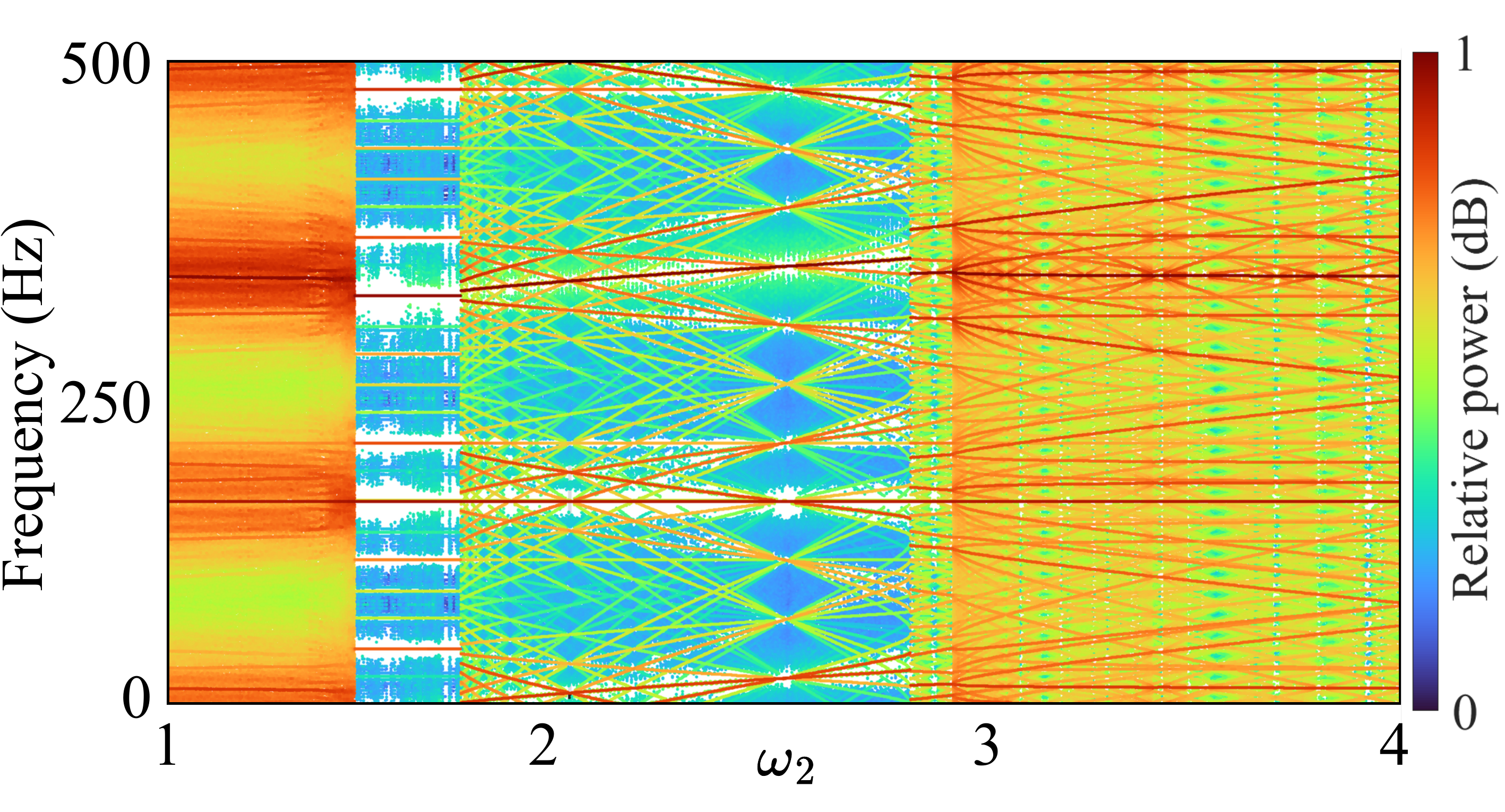}
			\caption{Numerical SBD for $\omega_2\in[1,4]$.}
			\label{tfqsbd_num}
		\end{subfigure}
		% Subfigure 2 
		\begin{subfigure}{\columnwidth}
			\centering
			\includegraphics[width=0.9\textwidth]{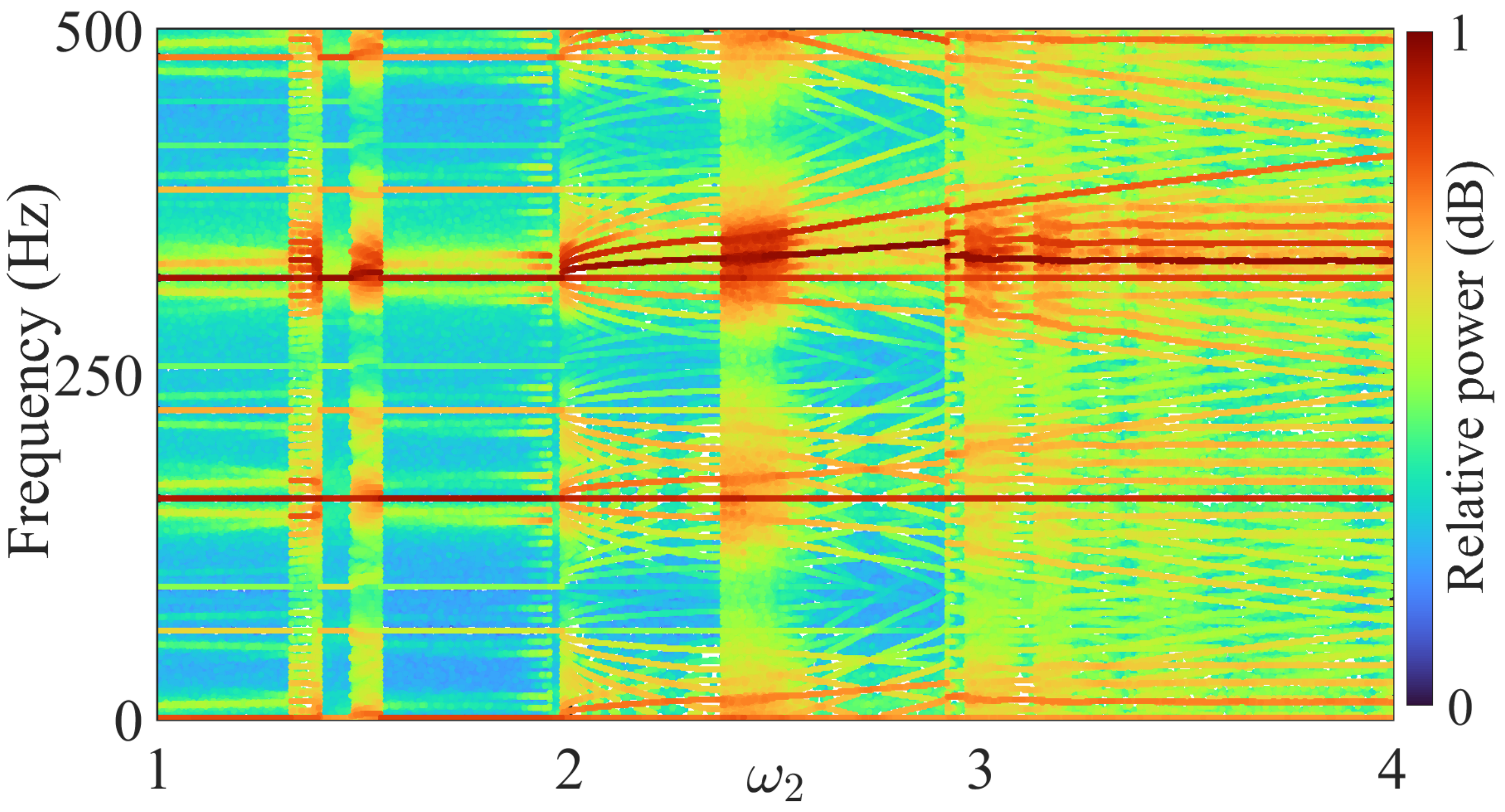}
			\caption{Experimental SBD for $\omega_2\in[1,4]$.}
			\label{tfqsbd_exp}
		\end{subfigure}
        % Subfigure 3 
        \begin{subfigure}{\columnwidth}
            \centering
		      \includegraphics[width =0.9\textwidth]{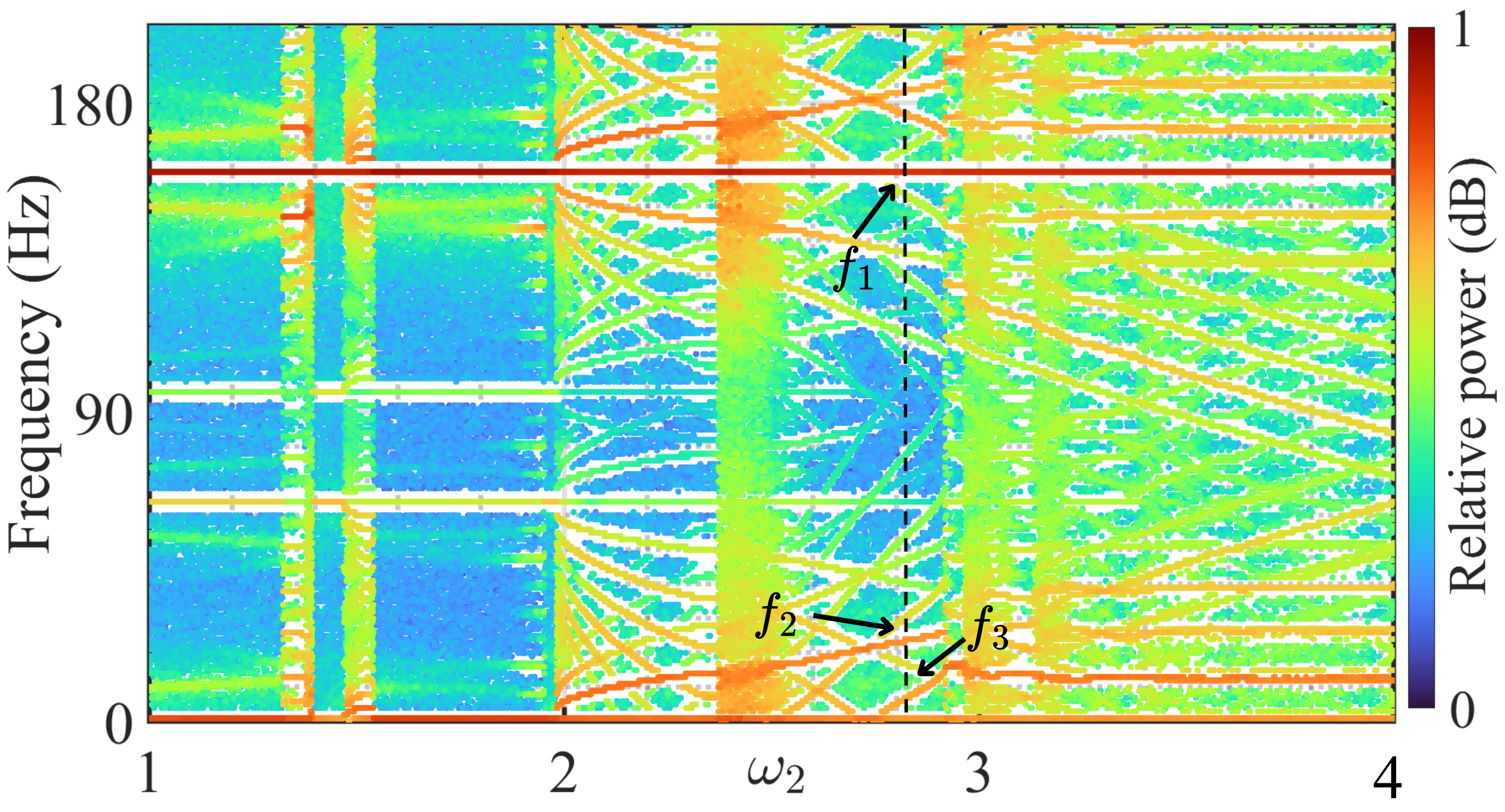}
		      \caption{Vertically zoomed SBD. $f_1,f_2,f_3$ mark the three incommensurate frequencies at $\omega_2=2.814$.}
		      \label{tfqsbd_expzoom}   
        \end{subfigure}
		\caption{Comparison of numerical and experimental SBDs for three frequency quasiperiodicity obtained from (\ref{batteqn}). Constant parameters: $\omega_1= 2.07,\beta = 0.097,\gamma = 2.52$.}
		\label{tfq_sbd}
	\end{figure}

    The numerically and experimentally obtained SBDs are shown in Fig.~\ref{tfq_sbd}. It is evident that many more fine structures are visible in the numerical SBD (Fig.~\ref{tfqsbd_num}) than in the experimental counterpart (Fig.~\ref{tfqsbd_exp}). The two SBDs also differ in exact parameter mapping; however, they both involve the interplay of three frequencies.

    The first frequency can be readily identified from Fig.~\ref{tfqsbd_exp} as $f_1=160~\text{Hz}$. To obtain the remaining two frequencies, we zoom in on the lowest rungs of the experimental SBD as in Fig.~\ref{tfqsbd_expzoom}. Now we look at the lowest branches, which appear to add up to form the spectrum. We note that in cases of messy spectra, one can get a better estimate of this by following the trick we discussed previously. This yields $f_2=23~\text{Hz}$ and $f_3=6~\text{Hz}$. This shows that the experimental SBD can be used to identify multiple incommensurate frequencies in the system.

\section{Conclusion} \label{conc}
In this work, we have demonstrated an automated method for generating spectral bifurcation diagrams (SBDs) and regular bifurcation diagrams for continuous-time nonlinear systems. In the process, we also provided the first experimental verification of the spectral signatures of various phenomena associated with Hopf bifurcations.

We illustrated the method by showing that the period-doubling bifurcations in the R\"ossler system appear as successive frequency halving in the experimental SBD. Although the experimental SBD is not devoid of experimental noise and parameter mismatches, the qualitative agreement with numerical results confirms the reliability of the proposed approach. 

In two-frequency quasiperiodicity (torus bifurcation), the experimental SBD confirms the earlier numerical observation about the occurrence of a `cross-hatching' pattern showing the interaction of two incommensurate frequencies. The experimental SBD also helps demarcate mode-locking regions in physically realized systems. 

The experimental SBD of the torus-doubling (double-covering) bifurcation also aligns with the numerical findings that one frequency halves while the other remains intact. 

We also observed three-frequency quasiperiodicity and captured its rich dynamics using SBD. 

These results demonstrate that spectral bifurcation diagrams are not just a numerical construct but a physically realizable and experimentally robust diagnostic tool. The study establishes SBD as a practical tool for analyzing nonlinear dynamical behavior directly from experimental data. Even in the presence of noise and other experimental errors, they capture the underlying frequency structure of nonlinear systems. 

The framework developed here can be extended to a wide range of real-world systems, including electronic, electrical, and mechanical systems, where identifying transitions between multiple frequencies is essential, thereby positioning SBDs as a promising experimental tool for analyzing complex dynamical behavior.

	\section*{Appendix}
	The codes for DAQ implementation, for Python-based parameter variation, and for generation of the experimental and numerical SBDs are available in the repository~\cite{code_repo}.

	\section*{Acknowledgment}
	
	The authors acknowledge the support from Dr. Soumyajit Seth for his guidance and thoughtful insights during experimental implementation. 

    \section*{Conflicts of interest}

    The authors declare that they have no known competing financial interests or personal relationships that could have appeared to influence the work reported in this paper
    
	\bibliographystyle{unsrt}  
\bibliography{Reference}

\end{document}